\documentclass{emulateapj}
\usepackage{apjfonts}

\usepackage{graphicx}
\usepackage{amsmath}
\usepackage{appendix}
\usepackage{amssymb}
\usepackage{aas_macros}
\usepackage{natbib}
\usepackage{latexsym}
\usepackage{longtable}
\usepackage{threeparttable}
\usepackage{subfigure}
\usepackage[figoff]{figcaps}
\usepackage{ccaption}
\captiondelim{ }

\begin{document}

% %%%%%%%%%%%%%%%%%%%%%%%%%%%%%%%%%%%
\pagestyle{plain}
\title{Orbits, masses, and evolution of main belt triple (87) Sylvia}
\author{Julia Fang\altaffilmark{1}, Jean-Luc Margot\altaffilmark{1,2}, Patricio Rojo\altaffilmark{3}}

\altaffiltext{1}{Department of Physics and Astronomy, University of California, Los Angeles, CA 90095, USA}
\altaffiltext{2}{Department of Earth and Space Sciences, University of California, Los Angeles, CA 90095, USA}
\altaffiltext{3}{Departamento de Astronomia, Universidad de Chile, Santiago, Chile}

\begin{abstract}

Sylvia is a triple asteroid system located in the main belt. We report
new adaptive optics observations of this system that extend the
baseline of existing astrometric observations to a decade.  We present
the first fully dynamical 3-body model for this system by fitting to
all available astrometric measurements.  This model simultaneously
fits for individual masses, orbits, and primary oblateness. We find
that Sylvia is composed of a dominant central mass surrounded by two
satellites orbiting at 706.5 $\pm$ 2.5 km and 1357 $\pm$ 4.0 km,
i.e.,\ about 5 and nearly 10 primary radii.  We derive individual
masses of 1.484$_{-0.014}^{+0.016}$ $\times 10^{19}$ kg for the
primary (corresponding to a density of 1.29 $\pm$ 0.39 g cm$^{-3}$),
7.33$_{-2.3}^{+4.7}$ $\times 10^{14}$ kg for the inner satellite, and
9.32$_{-8.3}^{+20.7}$ $\times 10^{14}$ kg for the outer satellite.
The oblateness of the primary induces substantial precession and the
$J_2$ value can be constrained to the range of 0.0985$-$0.1. The
orbits of the satellites are relatively circular with eccentricities
less than 0.04.  The spin axis of the primary body and the orbital
poles of both satellites are all aligned within about two degrees of each
other, indicating a nearly coplanar configuration and suggestive of
satellite formation in or near the equatorial plane of the primary.
We also investigate the past orbital evolution of the system by
simulating the effects of a recent passage through 3:1 mean-motion
eccentricity-type resonances.  In some scenarios this allow us to
place constraints on interior structure and past eccentricities.

\end{abstract}
\maketitle

% %%%%%%%%%%%%%%%%%%%%%%%%%%%%%%%%%%%
\section{Introduction} \label{introduction}

(87) Sylvia is a triple asteroid residing in the main belt, with
heliocentric semi-major axis 3.5 AU, eccentricity 0.085, and
inclination 11$^{\circ}$ relative to the ecliptic. Sylvia's outer
satellite, named Romulus, was discovered in 2001 using the W. M. Keck
Telescope \citep{brow01,marg01}, and was also detected in Hubble Space
Telescope (HST) images \citep{stor01}. The inner satellite, Remus, was
not discovered until the advent of improved adaptive optics systems in
2004 using the European Southern Observatory's Very Large Telescope
(VLT) \citep{marc05}. The diameter of the primary has been estimated
at $\sim$280 km through shape fits to adaptive optics images
\citep{marc05}; this estimate is consistent with stellar occultation
observations \citep{lin09}.  Assuming this primary size, approximate
sizes for the individual satellites have been estimated by adopting
the same albedo as the primary and measuring each satellite's
brightness relative to the primary.  The diameter
estimates are $\sim$7 km for Remus and $\sim$18 km for Romulus \citep{marc05}.

Sylvia was the first triple asteroid system announced, even though the
triple nature of 2002 CE26 was being actively discussed during the
acquisition of the Sylvia observations~\citep{shep06}.  Additional
discoveries of multiples in the Solar System have followed. They
include near-Earth triples (153591) 2001~SN263 \citep{nola08} and
(136617) 1994~CC \citep{broz09}, main belt triples Kleopatra
\citep{desc11}, Eugenia \citep{merl99,marc07}, Balam
\citep{merl02,marc08}, and Minerva \citep{marc11}, and trans-Neptunian
systems (47171) 1999~TC36 \citep{marg05,bene10}, Haumea
\citep{brow05,brow06}, and the Pluto/Charon system \citep{weav06}.

Following these discoveries, characterization of multiple systems have
unearthed a wealth of information about their fundamental physical
properties such as masses and densities, dynamical processes, and
constraints on formation and evolutionary mechanisms. Such research
has been possible because we can derive the masses of the individual
components of a triple or higher-multiplicity system by analyzing
their mutual gravitational interactions, which is possible in binary
systems only when reflex motion is detected~\citep{marg02s, ostr06,
naid11dps}. These masses in conjunction with size estimates can
provide densities. Using this method, \citet{fang11} performed a
detailed analysis of 2001~SN263 and 1994~CC, including masses,
densities, and dynamical evolution. Similarly, work on the
Pluto/Charon system and dwarf planet Haumea and its satellites have
yielded information about their physical properties, tidal
interactions, and evolutionary processes
\citep{lee06,thol08,rago09}. The high scientific return from studies
of binaries and triples has been reviewed by \citet{merl02} and
\citet{noll08}.

To date, no such dynamical orbit solution nor detailed analysis has
been performed for Sylvia. 
Previous work by \citet{marc05} approximated the actual orbits of
Remus and Romulus with individual two-body fits that included primary
oblateness. However, drawbacks of such methods include the failure to
account for third-body perturbations as well as the inability to solve
for individual component masses.  Additional researchers based their
studies on the published two-body orbits \citep{marc05} plus
unspecified component mass assumptions to study Sylvia's long-term
evolution \citep{wint09,frou11}, even though component masses are
undetermined and can span several orders of magnitude.

In this work, we report additional Keck and VLT imaging data for
Sylvia (Section \ref{observations}). Using primary$-$satellite
separations measured from these data plus published astrometry, we
present a fully dynamical 3-body orbital and mass solution for Sylvia,
by accounting for mutually interacting orbits as well as the primary's
non-sphericity (Section \ref{orbitsolution}). Although the orbital
periods of the satellites are near a 8:3 ratio, we do not find that
the system is currently in such a resonance (Section
\ref{currentmmr}).  
We also analyze Sylvia's short-term and long-term future evolution
(Section \ref{evolution}). Lastly, we investigate the past orbital
evolution of Remus and Romulus by modeling passage through the 3:1
mean-motion resonance (Section \ref{origin}). A summary of main
conclusions is given in Section \ref{conclusion}.

% %%%%%%%%%%%%%%%%%%%%%%%%%%%%%%%%%%%
\section{Observations} \label{observations}

We report new observations of Sylvia in 2011 from both Keck and VLT,
as well as summarize existing observations taken in 2001$-$2004 using
Keck, HST, and VLT. Astrometry derived from these datasets are used
for orbit fits described in the next section (Section
\ref{orbitsolution}).

\subsection{New Data in 2011}

Our observations in 2011 are summarized in Table \ref{obs}. In total,
we obtained 7 partial nights of service (or ``queue'') mode observing
at VLT and 4 partial nights of visitor mode at Keck.  At the VLT, we
used the NACO (NAOS-CONICA) high-resolution IR imaging camera 
\citep{rous03,lenz03} in its
S13 mode using the H filter, with a plate scale value reported as
0.013221 arcseconds per
pixel\footnote[1]{http://www.eso.org/sci/facilities/paranal/instruments/naco/doc}.
We used four offset positions in a box pattern, with an integration
time of 120 seconds per offset position. These four offset positions
sampled the four quadrants of the CCD to calibrate and mitigate
against detector defects. At Keck, we used NIRC2 (Near Infrared Camera
2) imaging in the H filter, with a plate scale of 0.009942 arcseconds
per
pixel\footnote[2]{http://www2.keck.hawaii.edu/inst/nirc2/genspecs.html}.
Our Keck observations used four offset positions in a similar box
pattern as with the VLT exposures, with an exposure time of 60 seconds
per offset position. All observations using both VLT and Keck were
performed with natural guide star adaptive optics, using Sylvia as the
guide star (its apparent magnitude varied from V$\sim$11.7 to
V$\sim$12.7 throughout the period of our 2011 observations).

\begin{figure}[htb]
	\centering
	\includegraphics[width=3.2in]{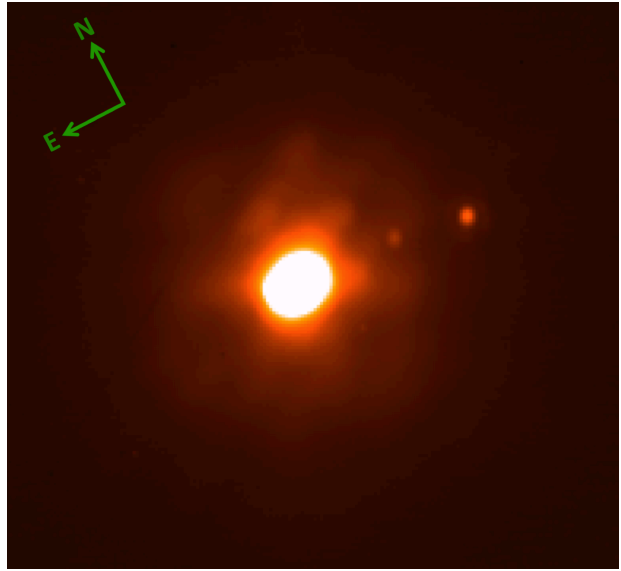}
	\caption{Keck H-band adaptive optics image on 2011 December 15
	(corresponding modified Julian Date 55910.2129) of Sylvia with
	inner satellite Remus and outer satellite Romulus. In this
	image, the primary$-$Remus separation is about 0.34
	arcseconds, and the primary$-$Romulus separation is about 0.57
	arcseconds.
\label{keckimage}} 
\end{figure}

We performed basic data reduction analysis for VLT and Keck
images. Each frame (at each dither position) is flat-fielded and its
bad pixels are corrected using a bad pixel mask obtained from outlier
pixels present in all frames. Sky subtraction is performed by
subtracting frames in pairs, where each frame is subtracted by another
frame where the target has been offset. We performed subpixel 2-D
Gaussian fitting to obtain precise centroids of the target in each
sky-subtracted frame, and these centroids were used to align and
combine all frames into one composite image. See Figure
\ref{keckimage} for an example of a composite image obtained using
Keck.

At each observation epoch, we detected either one or both satellites
(see Table \ref{obs}). As the satellites orbit the primary, they are
occasionally obscured by the bright primary and this occurs most often
for the inner satellite Remus. In cases where only one satellite is
detected, we determined the identification of the satellite through
orbit fitting. An incorrect identification of a satellite is easily
shown as an obvious outlier in orbit fits. In all cases examined here,
when only one satellite is detected, it is the outer satellite
Romulus.

For all satellite detections, we measure the astrometric positions of
the satellite relative to the primary by taking the difference between
the centroids of the primary and the satellite. Specifically, we
measure the position angle (degrees East of North) and separation of
the satellite relative to the primary. These measurements are
performed on the reduced composite image obtained at each observation
epoch.  These measurements are provided in Table \ref{astnew}.

% %%%%%%%%%%%%%%%%%%%%%%%%%%%%%%%%%%%
\def\arraystretch{1.4}
\begin{deluxetable*}{lcccl}
\tablecolumns{5}
\tablecaption{Summary of 2011 Observations \label{obs}}
\startdata
\hline \hline
\multicolumn{1}{c}{UT Date} &
\multicolumn{1}{c}{MJD} &
\multicolumn{1}{c}{Filter} &
\multicolumn{1}{c}{Telescope} &
\multicolumn{1}{c}{Detections} \\
\hline
2011 Oct 7	& 55841.1097	& H	& VLT	& Remus, Romulus \\
2011 Nov 6	& 55871.0856	& H	& VLT	& Romulus \\
2011 Nov 8	& 55873.1289	& H	& VLT	& Remus, Romulus \\
2011 Nov 10	& 55875.0451	& H	& VLT	& Remus, Romulus \\
2011 Nov 15	& 55880.0256	& H	& VLT	& Romulus \\
2011 Nov 16	& 55881.0466	& H	& VLT	& Remus, Romulus \\
2011 Nov 20	& 55885.0460	& H	& VLT	& Romulus \\
2011 Dec 15	& 55910.2129	& H	& Keck	& Remus, Romulus \\
2011 Dec 15	& 55910.2288	& H	& Keck	& Remus, Romulus \\
2011 Dec 15	& 55910.2698	& H	& Keck	& Remus, Romulus \\
2011 Dec 16	& 55911.1877	& H	& Keck	& Romulus \\
2011 Dec 16	& 55911.2510	& H	& Keck	& Romulus \\
2011 Dec 17	& 55912.2615	& H	& Keck	& Remus, Romulus \\
2011 Dec 18	& 55913.1972	& H	& Keck	& Romulus \\
2011 Dec 18	& 55913.2035	& H	& Keck	& Romulus
\enddata
\tablenotetext{}{Summary of our 2011 adaptive optics observations 
at VLT and Keck. Epochs are provided in Universal Time (UT) dates 
as well as the Modified Julian Date (MJD). Remus is the inner 
satellite and Romulus is the outer satellite.
}
\end{deluxetable*}
% %%%%%%%%%%%%%%%%%%%%%%%%%%%%%%%%%%%

% %%%%%%%%%%%%%%%%%%%%%%%%%%%%%%%%%%%
% %%%%%%%%%%%%%%%%%%%%%%%%%%%%%%%%%%%

% %%%%%%%%%%%%%%%%%%%%%%%%%%%%%%%%%%%
\def\arraystretch{1.4}
\begin{deluxetable*}{lrrrrrr}
\tablecolumns{7}
\tablecaption{Astrometry From 2011 Data \label{astnew}}
\startdata
\hline \hline
\multicolumn{1}{c}{Satellite} &
\multicolumn{1}{c}{MJD} &
\multicolumn{1}{c}{PA} &
\multicolumn{1}{c}{Sep.} &
\multicolumn{1}{c}{$x$} &
\multicolumn{1}{c}{$y$} &
\multicolumn{1}{c}{$\sigma$} \\
 &
 &
\multicolumn{1}{c}{(deg)} &
\multicolumn{1}{c}{(arcsec)} &
\multicolumn{1}{c}{(arcsec)} &
\multicolumn{1}{c}{(arcsec)} &
\multicolumn{1}{c}{(arcsec)} \\
\hline					
Remus   & 55841.1097 & 265.07 & 0.392 &  0.391 & -0.034 & 0.0132 \\
Remus   & 55873.1289 &  75.25 & 0.226 & -0.218 &  0.057 & 0.0132 \\
Remus   & 55875.0451 & 270.33 & 0.342 &  0.342 &  0.002 & 0.0132 \\
Remus   & 55881.0466 &  87.23 & 0.381 & -0.381 &  0.018 & 0.0132 \\
Remus   & 55910.2129 & 268.16 & 0.341 &  0.341 & -0.011 & 0.0099 \\
Remus   & 55910.2288 & 267.49 & 0.343 &  0.342 & -0.015 & 0.0099 \\
Remus   & 55910.2698 & 267.80 & 0.325 &  0.325 & -0.012 & 0.0099 \\
Remus   & 55912.2615 &  84.22 & 0.349 & -0.348 &  0.035 & 0.0099 \\
\hline
Romulus & 55841.1097 & 264.09 & 0.702 &  0.698 & -0.072 & 0.0132 \\
Romulus & 55871.0856 &  92.92 & 0.369 & -0.368 & -0.019 & 0.0132 \\
Romulus & 55873.1289 & 271.41 & 0.606 &  0.606 &  0.015 & 0.0132 \\
Romulus & 55875.0451 &  88.92 & 0.643 & -0.643 &  0.012 & 0.0132 \\
Romulus & 55880.0256 & 284.49 & 0.183 &  0.177 &  0.046 & 0.0132 \\
Romulus & 55881.0466 & 266.17 & 0.671 &  0.670 & -0.045 & 0.0132 \\
Romulus & 55885.0460 & 264.34 & 0.357 &  0.355 & -0.035 & 0.0132 \\                                        
Romulus & 55910.2129 & 264.30 & 0.571 &  0.568 & -0.057 & 0.0099 \\
Romulus & 55910.2288 & 263.66 & 0.572 &  0.568 & -0.063 & 0.0099 \\
Romulus & 55910.2698 & 265.11 & 0.541 &  0.539 & -0.046 & 0.0099 \\
Romulus & 55911.1877 &  95.41 & 0.392 & -0.391 & -0.037 & 0.0099 \\
Romulus & 55911.2510 &  92.09 & 0.446 & -0.446 & -0.016 & 0.0099 \\
Romulus & 55912.2615 &  78.76 & 0.423 & -0.415 &  0.082 & 0.0099 \\
Romulus & 55913.1972 & 272.80 & 0.528 &  0.528 &  0.026 & 0.0099 \\
Romulus & 55913.2035 & 272.09 & 0.521 &  0.521 &  0.019 & 0.0099
\enddata
\tablenotetext{}{Astrometry measured from 2011 data for Remus (inner)
and Romulus (outer) at specific MJD epochs. We measured the position
angle (PA; degrees East of North) and the separation of each satellite
relative to the primary. These values are converted to positions $x$
and $y$, where positive $x$ direction is towards the West, and
positive $y$ direction is towards the North.  We assign the instrument
plate scale as the positional uncertainty $\sigma$ for $x$ and $y$.  }
\end{deluxetable*}
% %%%%%%%%%%%%%%%%%%%%%%%%%%%%%%%%%%%

\subsection{Existing Data From 2001$-$2004}

Publicly-available astrometry datasets for Sylvia include Keck data in
2001 \citep{marg01}, HST data in 2001 \citep{stor01}, and VLT data in
2004 \citep{marc05}. For the 2004 VLT data, the paper by
\citet{marc05} contains their astrometric measurements of the
satellites relative to the primary expressed as $x-y$ pairs, where
they define $x$ and $y$ as positive in the East and North directions,
respectively.  However, they failed to indicate the signs (positive or
negative) for their $x$ and $y$ measurements of Remus. In addition, at
one epoch (MJD 53253.1738) their published astrometry give Remus an
implausibly large separation from the primary. To fix these
inaccuracies, we fit orbits to the astrometric points and have
determined the correct signs for their measurements (note that we
define $x$ to be positive in the West direction).  At epoch MJD
53253.1738, it appears that they have confused Remus and Romulus, and
so we have swapped measurements for these two bodies at that
epoch. These corrections, along with astrometry for the Keck and HST
data in 2001, are given in Table \ref{astold}.

In total, our baseline of observations spans a decade from 2001 to
2011. For Remus, which is never visible in data obtained prior to 2004, our
2011 observations add an additional 8 epochs to the existing 12 epochs
for a total of 20 epochs of astrometry, extending the baseline of
observations from about 1 month to 7 years.  For Romulus, our 2011
observations add an additional 15 epochs to the existing 30 epochs for
a total of 45 epochs of astrometric measurements, extending the
baseline of observations from 3 years to 10 years.  All of these
astrometry positions, given in Tables \ref{astnew} and \ref{astold},
are used for dynamical three-body orbit fits described in the next
section.

% %%%%%%%%%%%%%%%%%%%%%%%%%%%%%%%%%%%
\def\arraystretch{1.4}
\begin{deluxetable*}{lrrrrrrc}
\tablecolumns{8}
\tablecaption{Existing Astrometry From 2001$-$2004 Data \label{astold}}
\startdata
\hline \hline
\multicolumn{1}{c}{Satellite} &
\multicolumn{1}{c}{MJD} &
\multicolumn{1}{c}{PA} &
\multicolumn{1}{c}{Sep.} &
\multicolumn{1}{c}{$x$} &
\multicolumn{1}{c}{$y$} &
\multicolumn{1}{c}{$\sigma$} &
\multicolumn{1}{c}{Reference} \\
 &
 &
\multicolumn{1}{c}{(deg)} &
\multicolumn{1}{c}{(arcsec)} &
\multicolumn{1}{c}{(arcsec)} &
\multicolumn{1}{c}{(arcsec)} &
\multicolumn{1}{c}{(arcsec)} &
\multicolumn{1}{c}{} \\
\hline
Remus	& 53227.3042  &  ...  &  ...  & -0.411 &  0.002 & 0.0132 & 3 \\
Remus	& 53249.2470  &  ...  &  ...  & -0.239 & -0.101 & 0.0132 & 3 \\
Remus	& 53249.2532  &  ...  &  ...  & -0.228 & -0.101 & 0.0132 & 3 \\
Remus	& 53251.2985  &  ...  &  ...  &  0.246 &  0.107 & 0.0132 & 3 \\
Remus	& 53252.3627  &  ...  &  ...  &  0.421 &  0.000 & 0.0132 & 3 \\
Remus	& 53253.1738  &  ...  &  ...  & -0.445 & -0.025 & 0.0132 & 3 \\
Remus	& 53255.1091  &  ...  &  ...  &  0.435 &  0.002 & 0.0132 & 3 \\
Remus	& 53256.2886  &  ...  &  ...  &  0.268 & -0.072 & 0.0132 & 3 \\
Remus	& 53261.1432  &  ...  &  ...  & -0.394 &  0.033 & 0.0132 & 3 \\
Remus	& 53261.2298  &  ...  &  ...  & -0.430 & -0.008 & 0.0132 & 3 \\
Remus	& 53263.2146  &  ...  &  ...  &  0.412 &  0.004 & 0.0132 & 3 \\
Remus	& 53263.2202  &  ...  &  ...  &  0.416 &  0.001 & 0.0132 & 3 \\
\hline                 
Romulus & 51958.4810  & 97.00 & 0.564 & -0.559 & -0.069 & 0.0168 & 1 \\
Romulus & 51959.3660  & 60.30 & 0.425 & -0.369 &  0.211 & 0.0168 & 1 \\
Romulus & 51959.4200  & 54.10 & 0.383 & -0.310 &  0.225 & 0.0168 & 1 \\
Romulus & 51960.4010  &271.90 & 0.615 &  0.614 &  0.020 & 0.0168 & 1 \\
Romulus & 51962.4070  & 88.30 & 0.696 & -0.696 &  0.021 & 0.0168 & 1 \\
Romulus & 51963.5700  &306.00 & 0.330 &  0.267 &  0.194 & 0.0250 & 2 \\
Romulus & 53227.3042  &  ...  &  ...  & -0.377 &  0.144 & 0.0132 & 3 \\
Romulus & 53246.3105  &  ...  &  ...  & -0.785 & -0.097 & 0.0132 & 3 \\
Romulus & 53246.3659  &  ...  &  ...  & -0.755 & -0.117 & 0.0132 & 3 \\
Romulus & 53249.2470  &  ...  &  ...  & -0.547 &  0.139 & 0.0132 & 3 \\
Romulus & 53249.2532  &  ...  &  ...  & -0.555 &  0.136 & 0.0132 & 3 \\
Romulus & 53249.3516  &  ...  &  ...  & -0.654 &  0.109 & 0.0132 & 3 \\
Romulus & 53251.2985  &  ...  &  ...  &  0.763 & -0.058 & 0.0132 & 3 \\
Romulus & 53252.3627  &  ...  &  ...  &  0.156 &  0.214 & 0.0132 & 3 \\
Romulus & 53253.1738  &  ...  &  ...  & -0.791 &  0.049 & 0.0132 & 3 \\
Romulus & 53253.3445  &  ...  &  ...  & -0.834 & -0.018 & 0.0132 & 3 \\
Romulus & 53254.1603  &  ...  &  ...  & -0.172 & -0.214 & 0.0132 & 3 \\
Romulus & 53255.1091  &  ...  &  ...  &  0.835 &  0.003 & 0.0132 & 3 \\
Romulus & 53255.3928  &  ...  &  ...  &  0.793 &  0.105 & 0.0132 & 3 \\
Romulus & 53256.2886  &  ...  &  ...  & -0.272 &  0.202 & 0.0132 & 3 \\
Romulus & 53259.2030  &  ...  &  ...  &  0.683 &  0.165 & 0.0132 & 3 \\
Romulus & 53261.1432  &  ...  &  ...  & -0.546 & -0.186 & 0.0132 & 3 \\
Romulus & 53261.2298  &  ...  &  ...  & -0.449 & -0.205 & 0.0132 & 3 \\
Romulus & 53262.1602  &  ...  &  ...  &  0.724 & -0.073 & 0.0132 & 3 \\
Romulus & 53262.2759  &  ...  &  ...  &  0.786 & -0.035 & 0.0132 & 3 \\
Romulus & 53262.2815  &  ...  &  ...  &  0.791 & -0.032 & 0.0132 & 3 \\
Romulus & 53263.2146  &  ...  &  ...  &  0.221 &  0.230 & 0.0132 & 3 \\
Romulus & 53263.2202  &  ...  &  ...  &  0.215 &  0.231 & 0.0132 & 3 \\
Romulus & 53297.0193  &  ...  &  ...  & -0.752 & -0.038 & 0.0132 & 3 \\
Romulus & 53298.0064  &  ...  &  ...  &  0.101 & -0.186 & 0.0132 & 3

\enddata \tablenotetext{}{Existing astrometry measured from
2001$-$2004 data for Remus (inner) and Romulus (outer) at specific MJD
epochs, taken from [1] \citet{marg01}, [2] \citet{stor01}, and [3] \citet{marc05} 
(the latter with corrections; see text).  An ellipsis (...) means that
the value was not reported.  Measurements include the position angle
(PA; degrees Earth of North) and the separation of each satellite
relative to the primary. These values can be converted to positions
$x$ and $y$, where positive $x$ direction is towards the West, and
positive $y$ direction is towards the North. As in Table \ref{astnew},
we assign the instrument plate scale as the positional uncertainty
$\sigma$ for $x$ and $y$.  }
\end{deluxetable*}
% %%%%%%%%%%%%%%%%%%%%%%%%%%%%%%%%%%%

% %%%%%%%%%%%%%%%%%%%%%%%%%%%%%%%%%%%
\section{Orbital and Mass Solution} \label{orbitsolution}

We fit a fully dynamical 3-body model to the astrometric measurements
described in the previous section, taking into account mutually
interacting orbits.  Our model simultaneously fits for 16 parameters,
including a set of 6 orbital parameters per satellite (semi-major
axis, eccentricity, inclination, argument of pericenter, longitude of
the ascending node, and mean anomaly at epoch), 3 masses for the three
bodies, and the parameter $J_2$ representing the oblateness of the
primary.  The semi-major axis and eccentricity describe the size and
shape of the orbit, both the inclination and longitude of the
ascending node describe the orientation of the orbit, the argument of
pericenter describes the location of pericenter (minimum radial
distance of the orbit), and the mean anomaly at epoch can be used to
determine the location of the satellite in its orbit at a particular
time.

We describe these fitted parameters in more detail. The orbital
elements of each satellite are relative to the primary body, and are
defined with respect to the Earth equatorial reference frame of epoch
J2000.  Given that these orbital elements change over time due to
perturbations in the three-body system, these are defined as {\em
osculating} orbital elements. They are valid at a specific epoch, MJD
53227.0, corresponding to UT date 2004 August 10 00:00. The fitted 
masses are derived assuming $G$ = 6.67 $\times$ 10$^{-11}$ m$^3$ kg$^{-1}$ s$^{-2}$ 
for the gravitational constant. The primary's
non-spherical nature, which can introduce additional non-Keplerian
effects, is represented by an oblateness coefficient $J_2$. The
distribution of mass within the primary can be represented by terms in
a spherical harmonic expansion of its gravitational potential, and the
quadrupole term $J_2$ is the lowest-order gravitational moment. $J_2$
is related to the primary's three principal moments of inertia ($C
\geq B \geq A$) as
\begin{align} \label{j2equation}
	J_2 = \dfrac{C - \dfrac{1}{2}(A+B)}{MR^2},
\end{align}
where the denominator is a normalization factor including the
primary's mass $M$ and equatorial radius $R$
\citep[i.e.,][]{murr99}. In all of our fits, the radius $R$ is assumed
to be 140 km. The inclusion of $J_2$ in our fits also requires a
primary spin pole direction to be specified, and typically we fix the
primary pole to the orbit pole of the most massive satellite.

With a total of 16 parameters and 130 data measurements (Tables
\ref{astnew} and \ref{astold}), we have 114 degrees of freedom (number
of data points minus the number of parameters). We adopt a
least-squares approach to this problem by minimizing the chi-square
$\chi^{2}$ statistic, where $\chi^{2} = \displaystyle\sum\limits_{i}
(O_i-C_i)^2/\sigma_i^2$ for a set of $N (1 \leq i \leq N)$
observations with $\sigma_i$ uncertainties and observed $O_i$ and
computed $C_i$ values.  We utilize a Levenberg-Marquardt non-linear
least-squares algorithm written in IDL called {\verb mpfit }
\citep{mark09}. With 16 parameters, this is a very computationally
intensive problem, given the large amount of parameter space to
explore as well as the computationally expensive 3-body orbital
integrations that need to be performed. Especially for least-squares
problems with a large number of parameters, it is impossible to
guarantee that a global $\chi^{2}$ minimum has been found. More often
than not, the minimization procedure converged on a local minimum and
therefore it was necessary to re-fit with different starting
conditions. In total, we started the fitting procedure with tens of
thousands of sets of starting conditions, and we performed up to 20
iterations (equaling hundreds of $\chi^{2}$ evaluations) for each set
of starting conditions.

These starting conditions included plausible ranges of parameter
space for fitted parameters. We explored all possible values of
eccentricity (0$-$1), orbital angles (0$-$360$^{\circ}$), semi-major
axes (500$-$900 km for Remus and 1100$-$1500 km for Romulus), and
$J_2$ values (0$-$0.2). Starting values ranged from on the order of
10$^{18}$ to 10$^{20}$ kg for the primary's mass and from on the order
of 10$^{13}$ to 10$^{17}$ kg for each satellite's mass. Ranges for
satellite masses covered all possible mass values by sampling various
size \citep[Remus: $\sim$5$-$9 km in diameter, Romulus: $\sim$14$-$22
km in diameter; ][]{marc05} and density (0.1$-$10 g cm$^{-3}$) ranges.

In addition, we also explored various primary spin axis orientations
for the primary.  This was possible with this data set because the
perturbations due to the oblateness of the primary are detectable, and
because those perturbations depend on the spin axis orientation.
These effects are captured by three fitted parameters: two for the
primary spin axis orientation, and one for the value of $J_2$.  We
systematically explored the entire celestial sphere for the primary
spin axis orientation, but we also tested specific poles that had been
favored by previous studies.  These specific spin axis orientations
include RA=355$^{\circ}$ and DEC=82$^{\circ}$ (close to satellite
orbital poles) suggested from light curve analysis \citep{kaas02},
RA=68$^{\circ}$ and DEC=78$^{\circ}$ from a compilation of previous
data \citep{krys07}, and RA=100$^{\circ}$ and DEC=62$^{\circ}$ derived
using adaptive optics imaging data \citep{drum08}.  From these fits,
we find that primary spin poles misaligned with satellite orbit poles
do not provide good solutions (such as poles by \citet{krys07} and
\citet{drum08}).  Instead, we determined that the best-fit spin axis
direction was nearly aligned with the satellites' orbital poles, which
are almost coplanar.  As a result, for nearly all fits, we aligned the
primary's spin pole to the orbital pole of the most massive satellite.
The fact that both satellites orbit in or near the equator of the
primary provides an important constraint on satellite formation
mechanisms.

For each set of starting conditions, our orbit-fitting method
proceeded as follows. First, we performed N-body numerical
integrations using the {\verb Mercury } integration package
\citep{cham99}, which takes into account mutually interacting orbits
as well as the effects due to primary oblateness. We used a
Bulirsch-Stoer algorithm for our integration method, which is
computationally slow but accurate, and we chose an initial time step
that samples finer than 1/25th of the innermost orbital period. These
3-body integrations need to cover all epochs of observation, which
span about a decade (2001$-$2011). From these integrations, we
determined the positions and velocities of each satellite relative to
the primary at all epochs of observation (corrected for light travel
time) by interpolation. The length and resolution of these
integrations were the limiting factors in the computational speed of
each minimization.  Second, we obtained the vector orientation of
Sylvia's position relative to an observer on Earth for all epochs of
observation (again, corrected for light time), taking into account
aspect variations due to geocentric distance variations and Sylvia's
motion across the sky. Third, we used these orientations to project
and compute primary$-$satellite separations on the plane of the sky at
each observation epoch.  These computed separations were compared with
our observed separations (Tables \ref{astnew} and \ref{astold}) to
determine the $\chi^{2}$ goodness-of-fit statistic. These methods
benefit from the heritage of over a decade of work on orbit fitting as
well as our work on fitting 3-body models to datasets of triples,
including near-Earth asteroids \citep{fang11}.

\begin{figure}[htb]
	\centering \includegraphics[width=3.3in]{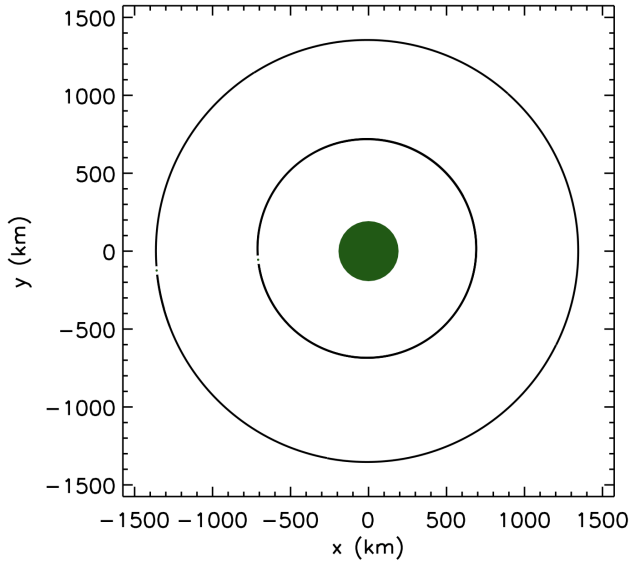}
	\caption{Diagram of best-fit orbits for satellites Remus
          (inner) and Romulus (outer), projected onto the primary's
          equatorial plane. These orbits show the actual trajectories
          from numerical integrations. The relative sizes of the
          bodies are shown to scale using green circles, assuming
          these diameters: 10.6 km for Remus, 10.8 km for Romulus, and
          280 km for the primary.  All bodies are located at their
          positions at MJD 53227 with the primary centered on the
          origin.
\label{orbitplot}} 
\end{figure}

\begin{figure}[htb]
	\centering
	\includegraphics[height=2.4in]{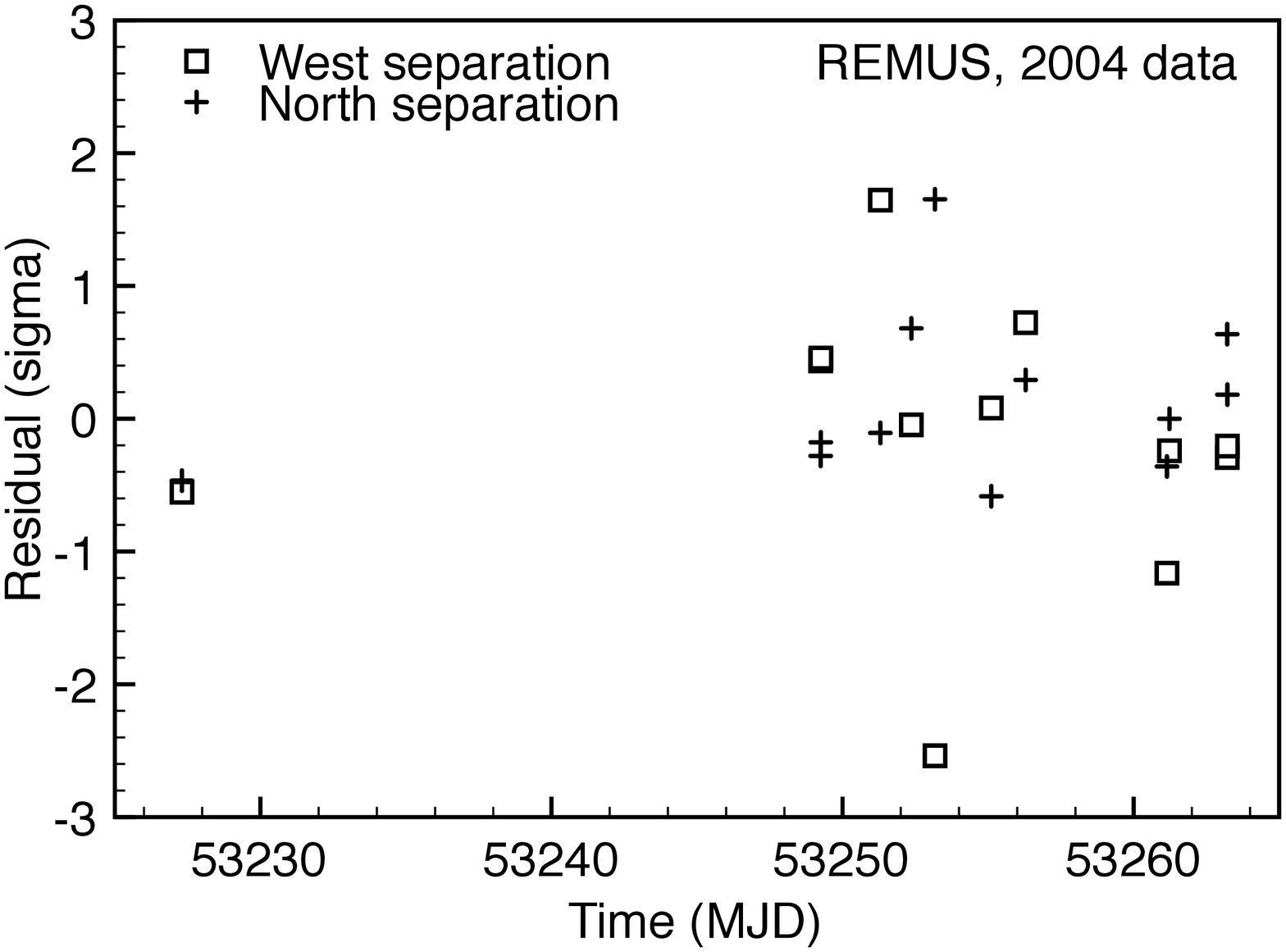}
	\includegraphics[height=2.4in]{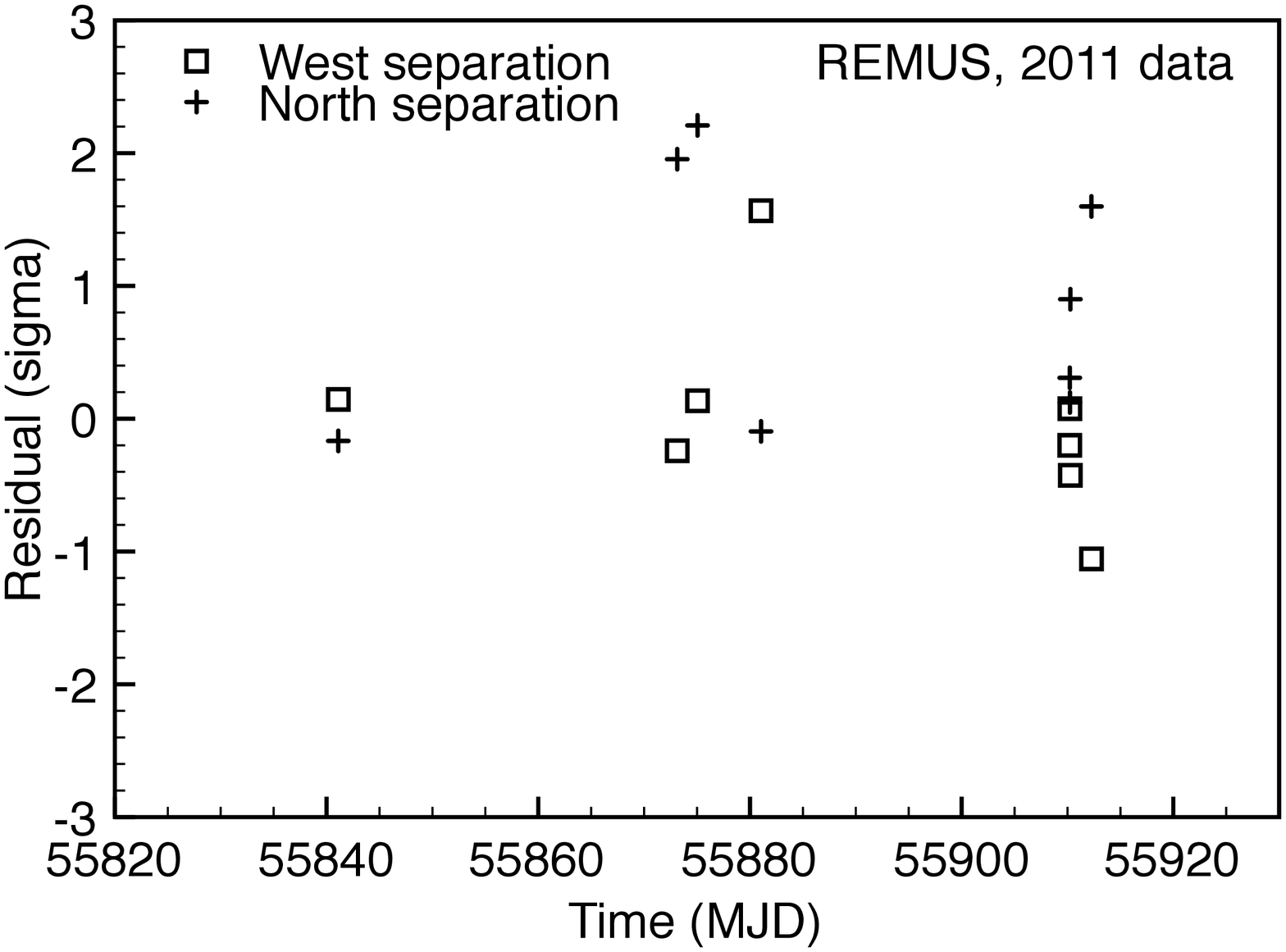}
	\caption{Residuals for separations in the West and North
          directions corresponding to the best-fit orbit (Table
          \ref{bestfit}) for inner satellite Remus. The two panels
          represent distinct epochs of observations (2004, 2011).
\label{remus_residuals}} 
\end{figure}

\begin{figure}[htb]
	\centering
	\includegraphics[height=2.4in]{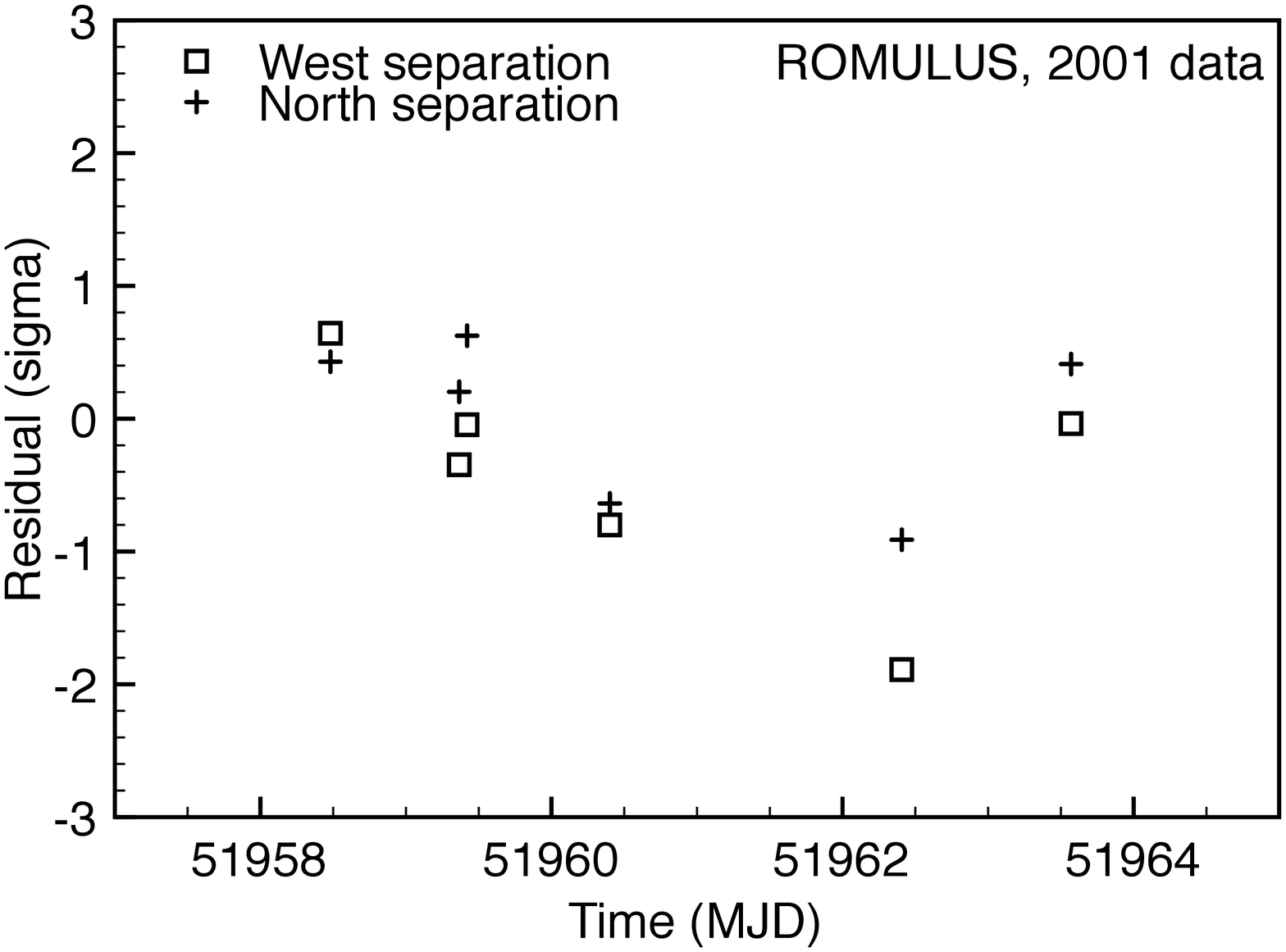}
	\includegraphics[height=2.4in]{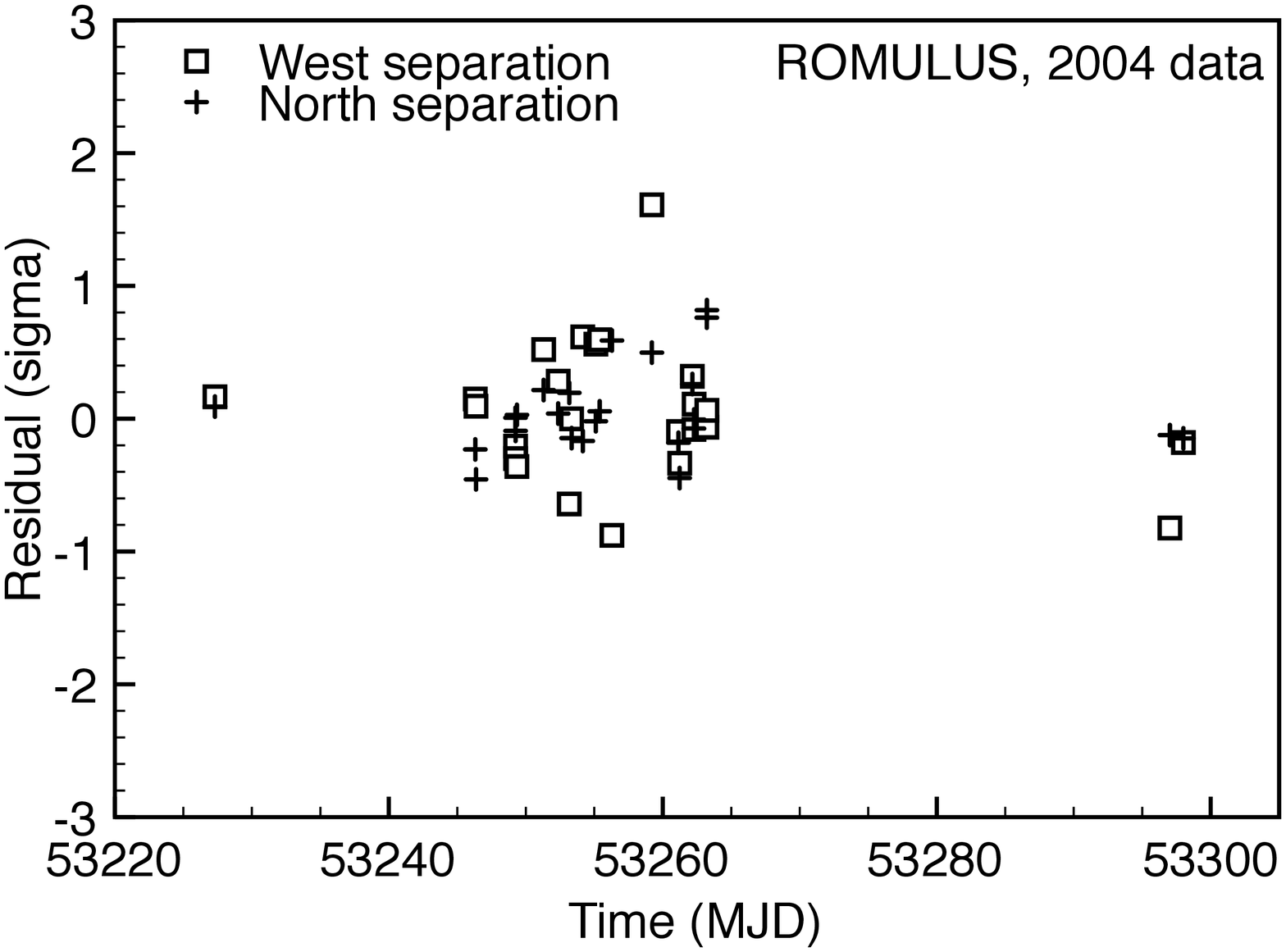}
	\includegraphics[height=2.4in]{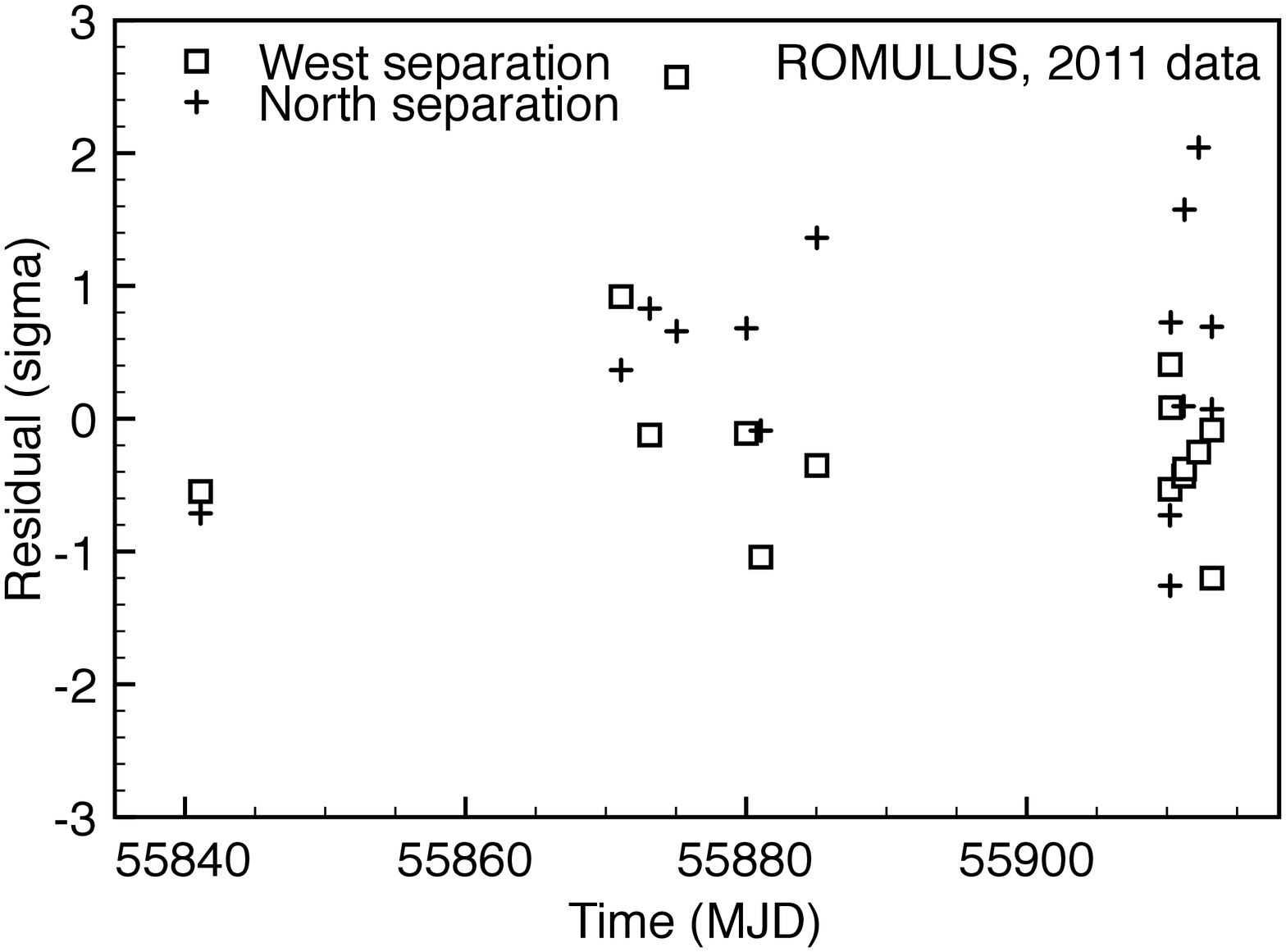}
	\caption{Residuals for separations in the West and North
          directions corresponding to the best-fit orbit (Table
          \ref{bestfit}) for outer satellite Romulus. The three panels
          represent distinct epochs of observations (2001, 2004, 2011).
\label{romulus_residuals}} 
\end{figure}

Table \ref{bestfit} shows the best-fit orbit solution.  The chi-square
is 73.55, and with 114 degrees of freedom, this corresponds to a
reduced chi-square of 0.6452.  This indicates that the fit is very
good, and that uncertainties were likely slightly overestimated.  This
best-fit solution is also visually illustrated in an orbit diagram in
Figure \ref{orbitplot}, where the orbits are projected onto the
primary's equatorial plane. Residuals of the best-fit solution are
shown in Figure \ref{remus_residuals} for Remus and Figure
\ref{romulus_residuals} for Romulus, where each figure's panel shows
the residuals for a particular year of observation (2001, 2004, or
2011). The residual is defined as $\chi_{i} = (O_i-C_i)/\sigma_i$ for
observation epoch $i$.

There are two types of 1$\sigma$ uncertainties given in Table
\ref{bestfit}: formal and adopted. The {\em formal} 1$\sigma$
uncertainties are obtained using the covariance matrix resulting from
the least-squares fitting procedure.  These formal 1$\sigma$
uncertainties may not always be accurate representations of the actual
uncertainties. Accordingly, the {\em adopted} 1$\sigma$ uncertainties
are obtained through a more rigorous method by determining each
parameter's 1$\sigma$ confidence levels
\citep[e.g.,][]{cash76,pres92}. To determine each parameter's
uncertainties, we hold the parameter fixed at a range of plausible
values while simultaneously fitting for all other parameters. Since
one parameter is held fixed at a time, a 1$\sigma$ confidence region
is prescribed by the range of solutions that yield chi-square values
within 1.0 of the lowest chi-square. This is a computationally
intensive process, and we performed this method to determine
uncertainties for the primary's pole (R.A. and Dec.) as well as for
each fitted parameter with the exception of the arguments of
pericenter and mean anomalies at epoch.  
We consider the adopted uncertainties to be more accurate
representations of the actual uncertainties than the formal
uncertainties. We note that these adopted uncertainties do not
preclude any systematic errors that may have occurred, such as during
the measurement of astrometry from the images.

Most solve-for parameters are well constrained by the data, with
formal uncertainties of 10\% or less, and as low as $\sim$1\% for the
mass and oblateness of the primary.  One notable exception is the mass
of Romulus.  The formal uncertainty on this parameter amounts to
$\sim$60\% of the nominal value, indicating that it is not well
constrained by the data.  As for orbit pole orientations, they are
determined to $\sim$1.5 degrees uncertainty (1$\sigma$) for Remus and
$\sim$1 degree for Romulus, and the primary spin axis orientation is
determined to $\sim$1 degree.  Given the near-circular nature of the
orbits, the arguments of pericenter $\omega$ and hence mean anomalies
at epoch $M$ are not well-defined (but the satellite positions, or
$\omega+M$, are in fact well-defined).

% %%%%%%%%%%%%%%%%%%%%%%%%%%%%%%%%%%%
\def\arraystretch{1.4}
\begin{deluxetable}{lrrr}
\tablecolumns{4}
\tablecaption{Best-fit Parameters and 1$\sigma$ Uncertainties \label{bestfit}}
\startdata
\hline \hline
\multicolumn{1}{l}{\bf Parameter} &
\multicolumn{1}{l}{\bf Best-fit} &
\multicolumn{1}{l}{\bf Formal 1$\sigma$} &
\multicolumn{1}{l}{\bf Adopted 1$\sigma$} \\
\hline
\multicolumn{4}{c}{\bf Remus (inner):} \\
Mass (10$^{14}$ kg) 	& 7.333   & $\pm$0.7172	    & $_{-2.333}^{+4.667}$ \\
$a$ (km) 		& 706.5   & $\pm$0.007231   & $_{-2.512}^{+2.488}$ \\
$e$ 			& 0.02721 & $\pm$0.009962   & $_{-0.01221}^{+0.01279}$ \\
$i$ (deg) 		& 7.824   & $\pm$0.6665     & $_{-0.8237}^{+0.6763}$ \\
$\omega$ (deg) 		& 357.0   & $\pm$15.14 	    & ...\\
$\Omega$ (deg) 		& 94.80	  & $\pm$5.000	    & $_{-5.802}^{+5.198}$ \\
$M$ (deg) 		& 261.0   & $\pm$13.43 	    & ...\\
$P$ (days) 		& 1.373   & ...             & $_{-0.009771}^{+0.01036}$ \\
\multicolumn{4}{c}{\bf Romulus (outer):} \\
Mass (10$^{14}$ kg) 	& 9.319    & $\pm$5.406	    & $_{-8.319}^{+20.68}$ \\
$a$ (km) 		& 1357 	   & $\pm$0.05918   & $_{-4.016}^{+3.984}$ \\
$e$ 			& 0.005566 & $\pm$0.004268  & $_{-0.003566}^{+0.005434}$ \\
$i$ (deg) 		& 8.293    & $\pm$0.2099    & $_{-0.2931}^{+0.2069}$ \\
$\omega$ (deg) 		& 61.06    & $\pm$18.74     & ...\\
$\Omega$ (deg) 		& 92.60	   & $\pm$1.339	    & $_{-1.597}^{+2.903}$ \\
$M$ (deg) 		& 197.0    & $\pm$18.75     & ...\\
$P$ (days) 		& 3.654    & ...            & $_{-0.02366}^{+0.02544}$ \\
\multicolumn{4}{c}{\bf Primary:} \\
Mass (10$^{19}$ kg) 	& 1.484     & $\pm$0.0001659 & $_{-0.01399}^{+0.01601}$ \\
$J_2$ 			& 0.09959   & $\pm$0.0008384 & $_{-0.001085}^{+0.0004148}$ \\
R.A. (deg) 		& 2.597     & $\pm$1.339     & $_{-1.597}^{+3.403}$ \\
Dec. (deg) 		& 81.71     & $\pm$0.2099    & $_{-0.7069}^{+0.2931}$
\enddata \tablenotetext{}{Best-fit parameters including individual
  masses, orbital parameters (semi-major axis $a$, eccentricity $e$,
  inclination $i$, argument of pericenter $\omega$, longitude of the
  ascending node $\Omega$, mean anomaly at epoch $M$), primary
  oblateness $J_2$, and primary spin pole (R.A. and Dec.).  These
  orbital elements are valid at epoch MJD 53227 in the equatorial
  frame of J2000. 
  We derived an effective orbital period $P$ from the best-fit values of
  semi-major axis and mass of the considered satellite plus all
  interior masses.  
  Two types of uncertainties are listed: formal 1$\sigma$ statistical
  errors are derived from the least-squares covariance matrix, and
  adopted 1$\sigma$ errors are obtained for select parameters through
  a more rigorous method (see text). For parameters $\omega$ and $M$
  with no adopted 1$\sigma$ errors, we recommend using the formal
  1$\sigma$ errors.
}
\end{deluxetable}
% %%%%%%%%%%%%%%%%%%%%%%%%%%%%%%%%%%%

% %%%%%%%%%%%%%%%%%%%%%%%%%%%%%%%%%%%
% %%%%%%%%%%%%%%%%%%%%%%%%%%%%%%%%%%%

% %%%%%%%%%%%%%%%%%%%%%%%%%%%%%%%%%%%
% %%%%%%%%%%%%%%%%%%%%%%%%%%%%%%%%%%%

% %%%%%%%%%%%%%%%%%%%%%%%%%%%%%%%%%%%
% %%%%%%%%%%%%%%%%%%%%%%%%%%%%%%%%%%%

The best-fit orbital solution indicates that the satellites follow
relatively circular orbits at semi-major axes of about 5 and nearly 10
primary radii. The mutual inclination (relative inclination) between
the orbital planes of the satellites is
0.56$^{\circ}$$_{-0.5}^{+1.5}$.
The low mutual inclination indicates that the orbital planes are
nearly coplanar.  This alignment supports a satellite formation
mechanism in which the satellites would tend to remain close to the
equatorial plane of the primary (e.g., a sub-catastrophic collision).
The alignment is likely indicative of formation conditions rather than
evolution, as tidal damping of inclinations can be a lengthy process.
Assuming various models (monolithic or rubble pile) for tidal
dissipation, inner satellite Remus could take on the order of 10$^8$
years up to the age of the Solar System to damp from 2 degrees to 1
degree.

Information about size, shape, and density can be derived from our
best-fit orbital solution.  We obtain a density of 1.29$\pm$0.39 g
cm$^{-3}$ for the primary by assuming a diameter of 280 km (lacking
realistic error bars on the size of the primary, we assumed volume
uncertainties of 30\%). The density error is dominated by the volume
error since the mass of the primary is known to $\sim$1\%.  The
primary is also oblate, with a well-constrained $J_2$ value of about
0.09959 which corresponds to an axial ratio $c/a=0.7086$ if we assume
equatorial symmetry and uniform density.  Size estimates for Remus and
Romulus can be obtained by assuming that they have a bulk density
equal to that of the primary, and by considering the adopted 1$\sigma$
confidence interval of satellite masses.  We find radii of
$\sim$4.5$-$6.1 km for Remus and $\sim$2.6$-$8.2 km for Romulus. These
ranges would have to be modified if the density of the primary or of
the satellites was different from the nominal value assumed here, but
only slightly as the dependence is $\rho^{-1/3}$.

We compare our best-fit solution in Table \ref{bestfit} to the
solution previously reported by \citet{marc05}. We find close
agreement in semi-major axes (within uncertainties).  The
eccentricities are marginally consistent.  Orbital plane orientations
differ by about $\lesssim$2 degrees.  The largest discrepancy between
our orbital solutions is the value of $J_2$. Our fits yield a very
well-constrained $J_2$ value with a 1$\sigma$ confidence range of
0.0985$-$0.1.  \citet{marc05} report two estimates for $J_2$ of $0.17 \pm
0.05$ and $0.18 \pm 0.01$, inconsistent with our value.
Using axial ratios from lightcurve analysis~\citep{kaas02} and a
uniform density assumption, we find $J_2=0.1$, in excellent agreement
with our dynamical value.

There are several possibilities to explain the discrepancies between
our orbital solutions and those of \citet{marc05}.  First, our orbital
fits are based on a much longer baseline of observations (2001$-$2011)
than their dataset (only 2004).  Second, our orbital solution is the
result of fully dynamical, N-body orbital fits that simultaneously
fitted for all parameters in the system, using numerical integrations
taking into account mutually interacting orbits and primary
oblateness. The fit obtained by \citet{marc05} used two-body
approximations (one satellite's orbit is fit at a time, ignoring
effects by the other satellite). These differences in dataset and
technique allowed us to obtain better-constrained orbital parameters
with smaller uncertainties as well as individual masses, which were
not previously known.

\section{Examination of Mean-Motion Resonance Occupancy} \label{currentmmr}

The orbital periods (ratio $\approx$ 2.661) of our best-fit solution
in Table \ref{bestfit} have a ratio near 8:3 (ratio $\approx$
2.667). To determine resonance occupation, we search for librating
resonance arguments using a general form of the resonance argument
\citep{murr99}
\begin{align} \label{mmr_argument}
	\phi = j_1\lambda_2 + j_2\lambda_1 + j_3\varpi_2 + j_4\varpi_1 + j_5\Omega_2 + j_6\Omega_1.
\end{align}
In Equation (\ref{mmr_argument}), $\phi$ is the resonant argument or
angle, $\lambda$ is the mean longitude, $\varpi$ is the longitude of
pericenter, and $\Omega$ is the longitude of the ascending
node. Subscripts 1 and 2 represent the inner and outer satellites,
respectively. The $j_i$ values (where $i = 1-6$) are integers and
their sum must equal zero (d'Alembert's rule). For the fifth-order 8:3
mean-motion resonance, $j_1 = -8$ and $j_2 = 3$ so we search through
integer values ($-30$ to $+30$) of the remaining $j_i$ values to
determine if there is libration of the resonant argument over
timescales ranging from 10 to 100 years.  To perform this search, we
determined the evolution of the relevant angles in Equation
(\ref{mmr_argument}) using 3-body numerical integrations with the
best-fit orbital, mass, and $J_2$ solution in Table \ref{bestfit}.  We
do not find any librating resonance arguments, and therefore we
conclude that the current system is not in the 8:3 mean-motion
resonance.

% %%%%%%%%%%%%%%%%%%%%%%%%%%%%%%%%%%%
\section{Short-term and Long-term Stability} \label{evolution}

In this section, we discuss the results of forward N-body integrations
of the best-fit orbital solution at MJD 53227 given in Table \ref{bestfit}. 
We perform short-term (50 yr) and long-term (1 Myr) simulations to 
determine how the orbital elements fluctuate with time and to 
assess the stability of this three-body system. Both short-term 
and long-term integrations are performed using a Bulirsch-Stoer 
algorithm in {\verb Mercury } \citep{cham99} with an initial timestep of 0.05 days, 
and include the 
gravitational effects of the three bodies and the primary's 
oblateness. For long-term integrations, we also include the effect of 
solar gravitational perturbations.

\begin{figure}[htb]
	\centering
	\includegraphics[width=3.3in]{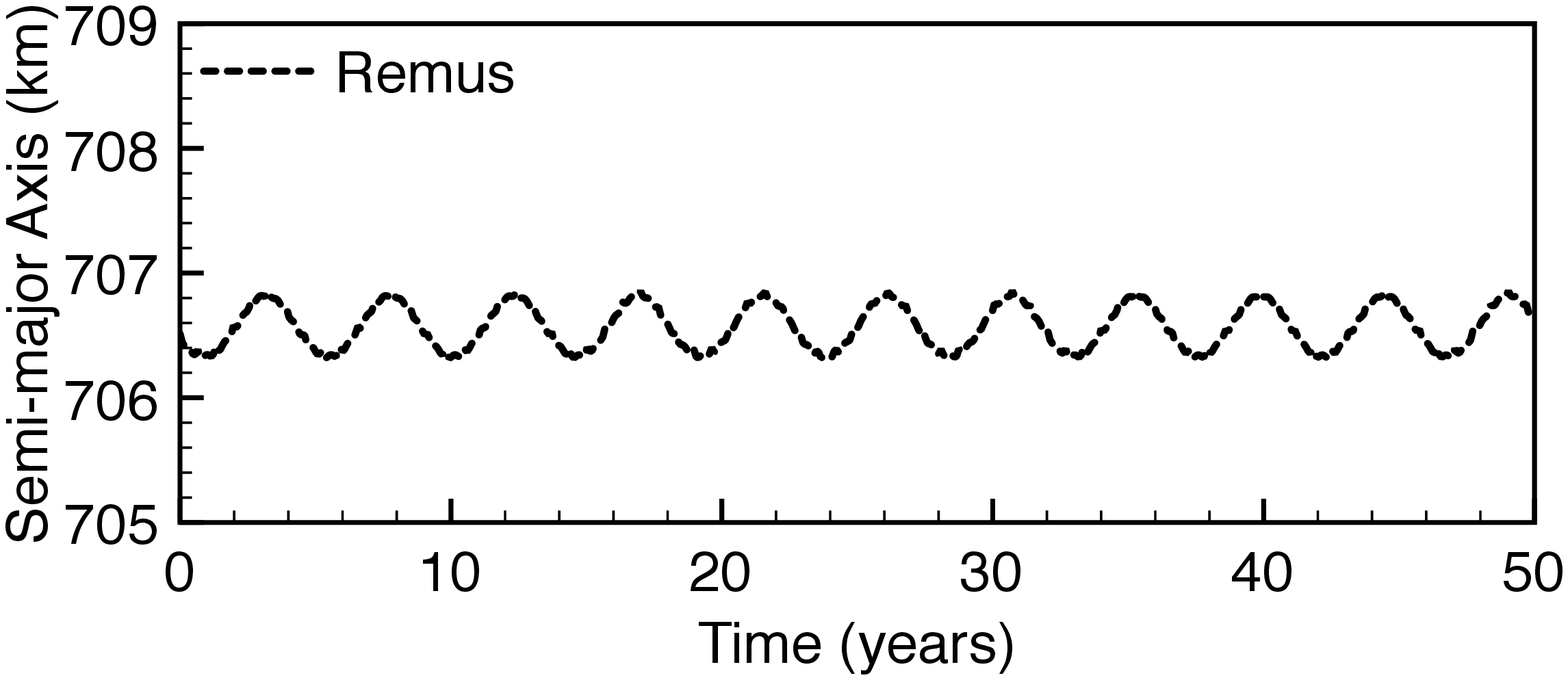}
	\includegraphics[width=3.3in]{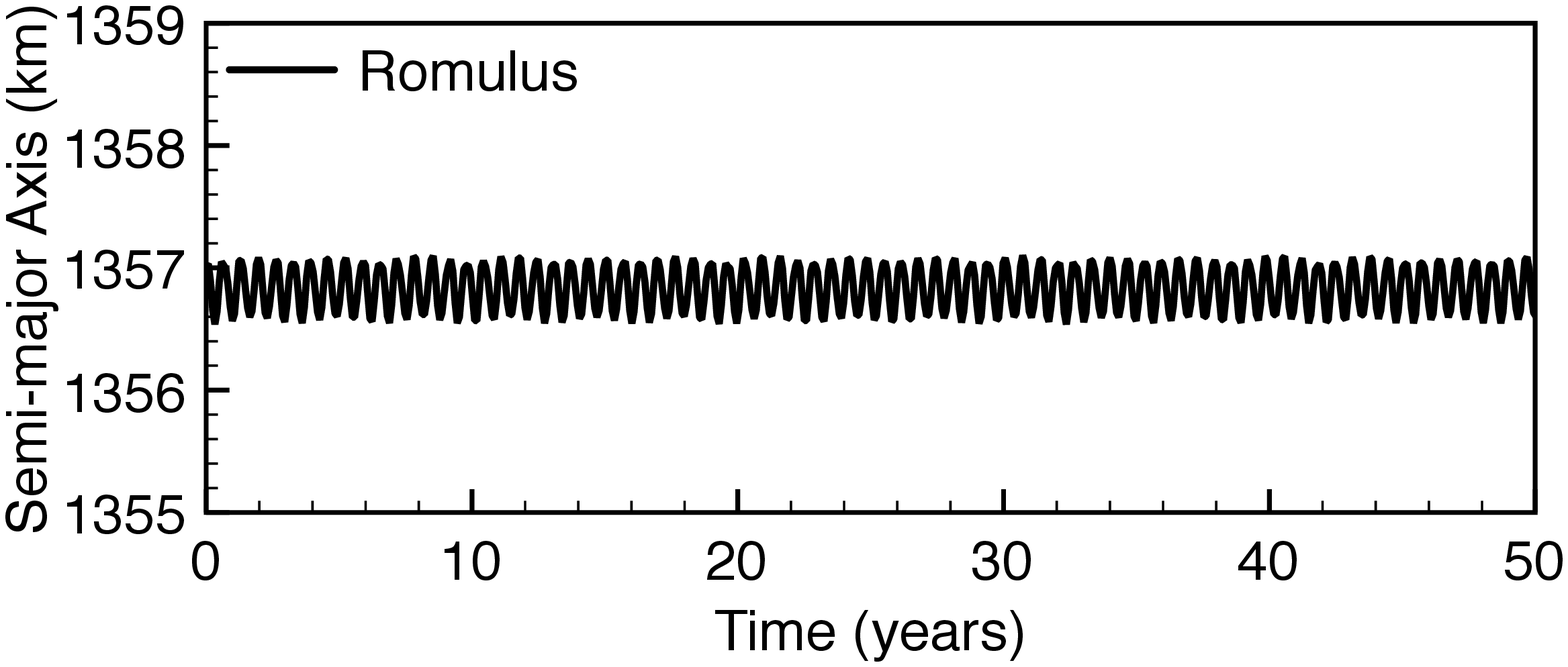}
	\includegraphics[width=3.3in]{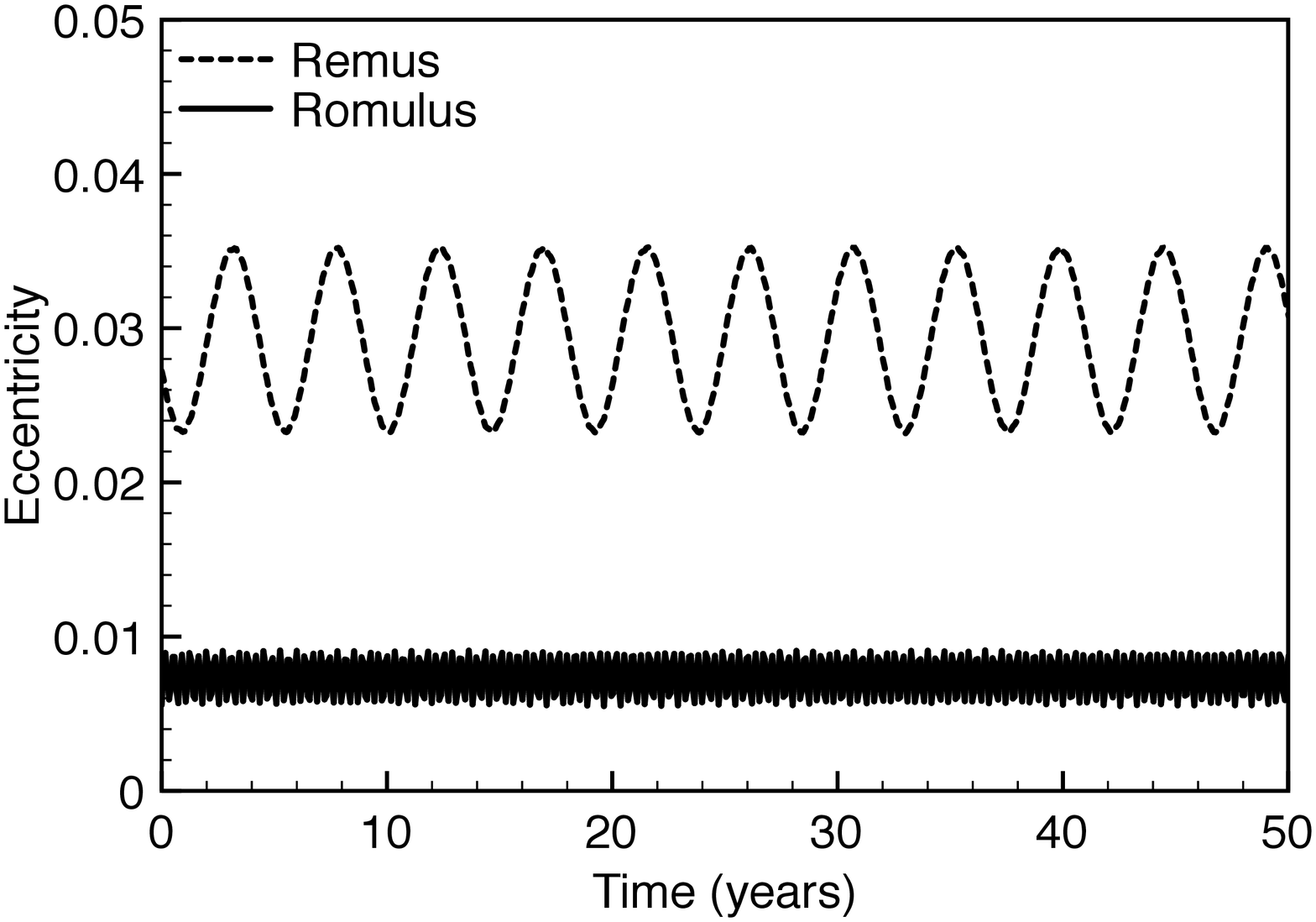}
	\caption{Variation of semi-major axis and eccentricity over a 50-year time span.
\label{mean_elements}} 
\end{figure}

The results of our short-term integrations are shown in Figure
\ref{mean_elements}. This figure illustrates how the semi-major axes
and eccentricities of both satellites evolve over a span of 50
years. From this figure, we can compute the mean value of orbital
elements, in contrast to the osculating orbital elements provided in
Table \ref{bestfit} that are valid at the specific epoch of MJD
53227.0. The semi-major axes for both Remus and Romulus have small
oscillations spanning less than 1 km, and have mean values of 706.57
km and 1356.83 km, respectively. The mean eccentricity values are
0.029 for Remus and 0.0074 for Romulus. The eccentricity variations
are especially apparent for Remus, whose eccentricity can vary from
0.023 to 0.035.  These short-period fluctuations are due to the effect
of the oblateness $J_2$ of the primary. The force due to the primary's
gravitational field can be modified to account for primary $J_2$, and
this modified force affects a satellite's orbit by inducing
short-period fluctuations in the semi-major axis, mean motion,
eccentricity, and mean anomaly.  Its effect on eccentricity can be
mathematically approximated as \citep{brou59,gree81}
\begin{align} \label{j2e}
	\Delta e \approx 3 J_2 (R_p/a)^2,
\end{align}
where $\Delta e$ is the maximum eccentricity excursion from minima to
maxima and $R_p$ is the primary's radius. Plugging in values of $J_2$
= 0.09959, $R_p$ = 140 km, and $a$ = 706.57 km (Remus) and 1356.83 km
(Romulus), we compute $\Delta e \approx 0.0117$ for Remus and $\Delta
e \approx 0.00318$ for Romulus. These values are consistent with the
maximum excursions seen in numerical simulations that are plotted in
Figure \ref{mean_elements}. Since Remus has a smaller separation
($\sim$5 $R_p$) from the primary than Romulus ($\sim$9.7 $R_p$), its
perturbation by primary $J_2$ is stronger, hence the larger
eccentricity variations seen in Figure \ref{mean_elements}.  As for
precession of the orbital planes, there is significant precession due to $J_2$ and
the presence of the other satellite. For example, Remus' longitude of
pericenter precesses $\sim$560$^{\circ}$ per year due to $J_2$ and
$\sim$1$^{\circ}$ per year due to Romulus.

% %%%%%%%%%%%%%%%%%%%%%%%%%%%%%%%%%%%
% %%%%%%%%%%%%%%%%%%%%%%%%%%%%%%%%%%%

The results from our long-term integrations show that this triple
system is stable over 1 Myr, and the variations in semi-major axis and
eccentricity do not noticeably exceed the fluctuations shown in Figure
\ref{mean_elements} for the short-term integrations. Accordingly, we
find that Sylvia is in a very stable configuration, as suggested by
its near circular and coplanar orbital state. We find that the
inclusion of the Sun's gravitational effect does not appreciably
affect the semi-major axes and eccentricities of Sylvia's satellite
orbits.  If we consider Sylvia's Hill sphere of gravitational
influence, defined as $r_{\rm Hill} = a_{\odot}(M/(3M_{\odot}))^{1/3}$
($a_{\odot}$ is the heliocentric semi-major axis, $M$ is the primary's
mass, and $M_{\odot}$ is the mass of the Sun), we calculate that the
satellites orbit at $\sim$1\% and $\sim$2\% of the Hill radius. As a
result, they are both well within the primary's sphere of
gravitational dominance over the Sun.

Our stability results are in agreement with two previous
investigations on Sylvia's stability. \citet{wint09} performed
stability analyses of the system, including the effects of the Sun and
Jupiter. They find that Sylvia is not stable unless the primary has at
least a minimal amount of oblateness (0.1\% of their assumed primary
$J_2$ of 0.17).  They show that the inclusion of primary oblateness
gives rise to a secular eigenfrequency that is much faster than those
induced by other gravitational perturbations, which provides a
stabilizing effect on the satellites' orbital evolution. 
\citet{frou11} investigated Sylvia's short-term (20 years) and
long-term evolution (6600 years) including the primary's
non-sphericity (assuming $J_2 \approx 0.14$) and solar
perturbations. They also varied the semi-major axis and eccentricity
of the satellites' orbits to determine the extent of their stability
zones. They find that the current configuration of the system lies in
a very stable zone. Authors from both papers \citep{wint09,frou11} 
mention that the effect of Jupiter is negligible compared 
to the effect of the Sun.

% %%%%%%%%%%%%%%%%%%%%%%%%%%%%%%%%%%%
\section{Evolution of Orbital Configuration} \label{origin}

In this section, we investigate the past orbital evolution of Remus
and Romulus.  We find that tidal perturbations can cause the orbits to
evolve and to cross mean-motion resonances.  This resonance passage
may perturb orbits by increasing eccentricities.  First we discuss how
tidal processes likely caused the satellites to encounter the 3:1
mean-motion resonance in their past, then we describe our numerical
modeling methods, and lastly we present plausible past evolutionary
pathways as suggested by our simulation results.

\subsection{Tidal Theory}

Tidal evolution can cause the semi-major axis of an orbit to expand
due to tides raised on the primary by its satellite
\citep{gold63,gold66}. Tides raised on the satellite by the primary
have an insignificant effect on the semi-major axis. The rate of
semi-major axis evolution is given as
\begin{align} \label{dadt}
	\dfrac{da}{dt} = 3 \dfrac{k_p}{Q_p}\dfrac{M_s}{M_p}\left(\dfrac{R_p}{a}\right)^{5}na, 
\end{align}
where $a$ is the semi-major axis, $k$ is the tidal Love number, $Q$ is
the tidal dissipation factor, $M$ is the mass, $R$ is the radius, and
$n$ is the mean motion. The subscripts $p$ and $s$ denote quantities
for the primary and satellite, respectively. It is likely that tidal
evolution is causing the orbits of Remus and Romulus to expand, and
that their orbits were in a more compact configuration in the past. We
discuss the relative importance of tides compared to another important
evolutionary process (BYORP) at the end of this subsection.

We expect that orbital expansion by tides is causing the relative
orbits of Remus and Romulus to slowly converge towards each other. Two
orbits are converging if $\dot{a}_{\rm 1}/\dot{a}_{\rm 2}$ is greater
than one, and here subscripts $1$ and $2$ represent Remus (inner) and
Romulus (outer), respectively. Using Equation (\ref{dadt}), we can
express this criterion as
\begin{align} \label{converging}
	\dfrac{\dot{a}_{\rm 1}}{\dot{a}_{\rm 2}} = \dfrac{M_1}{M_2}\left(\dfrac{a_2}{a_1}\right)^{11/2} > 1.
\end{align}
Assuming best-fit values for the semi-major axes of Remus and Romulus
(Table \ref{bestfit}), their orbits are currently converging as long
as their mass ratio satisfies $M_1/M_2 > 0.0276$. Taking into account
the range of the 1$\sigma$ adopted confidence interval in masses (from
Table \ref{bestfit}), we find that this ratio is satisfied in all
cases and therefore we expect that their orbits are converging.  The
steep dependence of tidal evolution on semi-major axis causes the
orbit of Remus to expand much faster than the orbit of Romulus. Given
that their orbits are slowly converging over time, we can determine
the most recent mean-motion resonance passage encountered by the
satellites.  By considering all first, second, third, and fourth order
resonances (where a $p+q$:$p$ resonance is $q$th order), we expect
that the most recent resonance encountered by the system is the
second-order 3:1 resonance. Accordingly, in our analysis here we focus
on the 3:1 resonance and its effect on the satellites' eccentricity
evolution.

We describe how tidal evolution causes the eccentricity of an orbit to
increase or decrease. This is a competing process between the opposing
effects of tides raised on the primary (eccentricity increases) and
tides raised on the satellite (eccentricity decreases). These two
opposing effects are contained in two terms in the equation
\citep{gold63,gold66}
\begin{align} \label{dedt}
	\dfrac{de}{dt} = \dfrac{57}{8}\dfrac{k_p}{Q_p}\dfrac{M_s}{M_p}\left(\dfrac{R_p}{a}\right)^5ne
	- \dfrac{21}{2}\dfrac{k_s}{Q_s}\dfrac{M_p}{M_s}\left(\dfrac{R_s}{a}\right)^5ne,
\end{align}
which gives the evolution of eccentricity $e$. The variables and
subscripts in Equation (\ref{dedt}) are the same as for Equation
(\ref{dadt}).

\begin{figure}[thbp]
	\centering
	\includegraphics[width=3.3in]{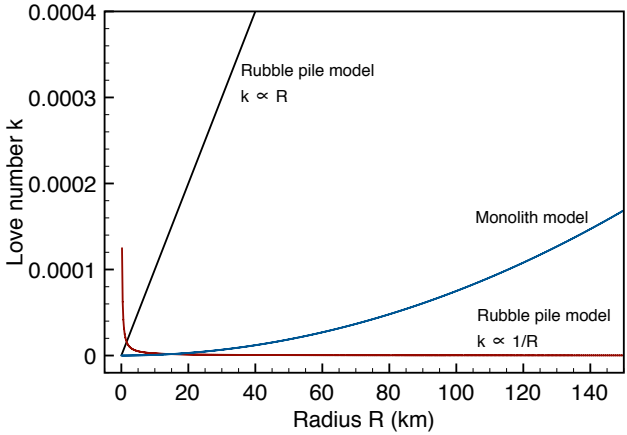}
	\caption{Tidal Love number models (Equations
	\ref{monogold}$-$\ref{rubbjaco}) as a function of
	radius. Intersections between the $k \propto 1/R$ model and
	the other two models occur at 1.58 km and 14.94 km. 
 \label{loveplot}}
\end{figure}

We discuss models for the tidal Love number $k$ in Equations
(\ref{dadt}) and (\ref{dedt}) used for calculations of tidal evolution
in asteroids. These models are dependent on the asteroid's radius
$R$. First, we consider the monolith model where
\begin{align} \label{monogold}
	k \approx \dfrac{1.5}{1 + 2 \times 10^8\ \left(\dfrac{1\ \rm km}{R}\right)^2},
\end{align}
and this model is appropriate for asteroids that are idealized as
uniform bodies with no voids \citep{gold09}. Second, we consider a
rubble pile model by \citet{gold09} with the following Love number
formalism:
\begin{align} \label{rubbgold}
	k \approx 1 \times 10^{-5} \left(\dfrac{R}{1\ \rm km}\right).
\end{align}
Rubble pile models are appropriate for asteroids idealized as
gravitational aggregates. Another rubble pile model is given by
\citet{jaco11} where
\begin{align} \label{rubbjaco}
	k \approx 2.5 \times 10^{-5}\ \left(\dfrac{1\ \rm km}{R}\right). 
\end{align}
Equation (\ref{rubbjaco}) was obtained by fitting to the
configurations of known asteroid binaries and assuming they are in an
equilibrium state with tidal and BYORP effects canceling each other.
Comparison between these three Love number models are given in Figure
\ref{loveplot}.

% %%%%%%%%%%%%%%%%%%%%%%%%%%%%%%%%%%%
% %%%%%%%%%%%%%%%%%%%%%%%%%%%%%%%%%%%

Here we assume that tidal evolution is a dominant process and we do
not analyze other evolutionary processes such as BYORP
perturbations. BYORP is a radiative effect that is predicted to cause
orbital evolution of a synchronous satellite on short timescales
\citep{cuk05,cuk07}, and has not been observationally verified yet.
BYORP effects dominate at larger semi-major axes, and tides dominate
at smaller semi-major axes \citep{jaco11}. 
Tides result in an increase of the semi-major axes, but BYORP can
result in an increase or a decrease, depending on the shapes of the
satellites, which are unknown.  For the purpose of our evolution
calculations,
the relevant semi-major axis rate is the {\em relative} migration rate
of the satellites, and we compare the contributions by tides and BYORP
by calculating $|(\dot{a}_{\rm tides,1}-\dot{a}_{\rm
  tides,2})/(\dot{a}_{\rm byorp,1}-\dot{a}_{\rm byorp,2})|$, with
subscripts $1$$=$inner and $2$$=$outer.  We find that this quantity is
$\sim$1.4$-$3.3 (taking into account whether BYORP acts in the same or
opposite directions for both satellites), and therefore we expect that
tidal evolution will dominate the relative 
rate of the satellites' orbits as they converge.  This calculation
assumes current semi-major axes, $Q_p=100$, and a monolith tidal Love
number model for the primary.  If BYORP 
causes the orbits of the satellites to converge, then joint evolution
by tides and BYORP will result in a higher relative migration rate
than the tides-only evolution considered here. If BYORP works against
tides, then their joint evolution will be slower. Given the
order-of-magnitude uncertainties already inherent in unknown tidal
parameters such as $Q$ and $k$ as well as uncertainties in whether
BYORP will expand or contract the satellites' orbits, we do not
include the effects of BYORP in our simulations.

\subsection{3:1 Eccentricity-type Resonances}

The 3:1 mean-motion resonance is the most recent low-order resonance
encountered by Remus and Romulus, and here we briefly describe the
relevant eccentricity-type resonances that can affect orbital
eccentricities. We do not consider 3:1 inclination-type resonances in 
our simulations.  There are 3 eccentricity-type
resonances for the 3:1 mean-motion resonance: $e_2^2$ resonance which
perturbs only the outer satellite's eccentricity, $e_1e_2$ mixed
resonance which perturbs both satellites' eccentricities, and the
$e_1^2$ resonance which perturbs only the inner satellite's
eccentricity. Subscripts $1$ and $2$ represent Remus (inner) and
Romulus (outer), respectively.

The relevant resonance arguments $\phi$ for these three 3:1
eccentricity resonances are \citep[e.g.,][]{murr99}
\begin{align} \label{arg1}
	e_2^2:  \hspace{20 mm} \phi = 3\lambda_2 - \lambda_1 - 2\varpi_2,
\end{align}
\begin{align} \label{arg2}
	e_1e_2: \hspace{13 mm} \phi = 3\lambda_2 - \lambda_1 - \varpi_1 - \varpi_2,
\end{align}
\begin{align} \label{arg3}
	e_1^2:  \hspace{20 mm} \phi = 3\lambda_2 - \lambda_1 - 2\varpi_1,
\end{align}
where $\lambda$ is the mean longitude and $\varpi$ is the longitude of
pericenter. Occupation of any of these resonances requires libration
of the considered resonance argument (exact resonance occurs when
$\dot{\phi}=0$). Note that $\dot{\lambda}=n$, where $n$ is the mean
motion and is related to the semi-major axis.

The resonance arguments given in Equations (\ref{arg1})$-$(\ref{arg3})
are listed in the order that these resonances are encountered due to
tidal migration: first the $e_2^2$ (at $a_1/a_2 \approx 0.481$), then
the $e_1e_2$ (at $a_1/a_2 \approx 0.483$), and lastly the $e_1^2$ (at
$a_1/a_2 \approx 0.485$), where the resonance locations $a_1/a_2$ must
be adjusted depending on the exact starting values of $a_1$ and $a_2$.
These resonances are not located at the same semi-major axes;
differentiation of Equations (\ref{arg1})$-$(\ref{arg3}) shows that
the various resonant arguments will librate at different values of
$n_1$ and $n_2$. Such ``resonance splitting'' occurs because
perturbations such as the effect of primary $J_2$, and to a lesser
degree (in this case, 2$-$3 orders of magnitude smaller), secular
perturbations, causes $\varpi$ of the satellites to precess at
different rates.  In Section \ref{mmr_results}, we will discuss the
capture of Remus and Romulus into any of these resonances. Next, we
describe our methods regarding the implementation of tidal effects
using direct N-body integrations.

\subsection{Methods}

Our methods and implementation for simulating a 3:1 resonant passage
due to tidal migration are as follows. We use an N-body integrator
with a variable-timestep Bulirsch-Stoer algorithm from 
{\verb Mercury } \citep{cham99}. 
We implement additional terms in the equations of motion due to the
effects of tides on semi-major axis and eccentricity by following the
numerical methods described in Appendix A of \citet{lee02}.
Specifically, we used Equations (\ref{dadt}) and (\ref{dedt}) to model
the tidal evolution in time of $a$ and $e$.  We have tested our
implementation by reproducing results in \citet{lee02} as well as
matching the analytical expectations (Equations (\ref{dadt}) and
(\ref{dedt})) of semi-major axis drift and eccentricity evolution
outside of resonance.

Actual tidal timescales can be computationally prohibitive, and we
incorporate a ``speedup'' factor to artificially increase the rate of
tidal evolution in our simulations. Such speedup factors have also
been numerically implemented in previous studies of tidal migration
\citep[i.e.,][]{ferr03,meye08,zhan09}, where they adopted values up to
1000 and found that their results were not sensitive to the choice of
speedup factor in the range 1-1000.  In our implementation, we
incorporate a speedup factor by multiplying Equations (\ref{dadt}) and
(\ref{dedt}) by typical speedup factors of $100-1000$. In agreement
with previous studies, we find that our results are not sensitive to
the choice of speedup factors up to 1000 for select test cases.

We integrate the system for artificial durations of 1-10 Myr, which,
because of the speedup factors, represent 1 Gyr of tidal evolution.
Our figures show the tidal evolution timescale, not the artifical
timescale used in the integrator.
The 1 Gyr timescale is constrained by the lifetime of Sylvia's
satellite system. Work by \citet{vokr10} investigating the
collisionally-born asteroid family related to Sylvia suggests that the
family members (and hence the satellites) are at least 10$^8$ years
old. We can also estimate the lifetime of the satellites by
considering how much time would pass before a collision between one of
the satellites and another main belt asteroid. We estimate this
timescale to be roughly 10$^9$ years \citep{fari98,bott05} by assuming
that the smallest satellite has a diameter of 10.6 km, as suggested by
our orbital fit analysis (Section \ref{orbitsolution}). Accordingly,
we consider 1 Gyr to be a reasonable time within which tidal evolution
can have taken place, and we typically do not run simulations longer
than 1 Gyr.

For the default set of initial conditions in our simulations, the
masses of all bodies and primary oblateness ($J_2$) are taken from
Table \ref{bestfit}. Simulations are started with coplanar and nearly
circular ($e = 0.001$) orbits. Angles for the argument of pericenter
and mean anomaly are given a random value from 0$^{\circ}$ to
360$^{\circ}$. Initial semi-major axes for Remus and Romulus are 654
km and 1352.5~km, just inside the 3:1 resonance location. To simulate
tidal evolution, we also need to adopt values for the Love number $k$
and tidal dissipation factor $Q$. To calculate the Love number, we use
the tidal monolith model for these bodies \citep[see
e.g.,][]{gold09}. For all bodies, we assume $Q = 100$, a reasonable
assumption for rocky monoliths.

We also ran additional sets of simulations where we varied the initial
eccentricities of both satellites (0.001$-$0.050), primary $J_2$
(5$-$10\% lower and higher than its best-fit value), satellite masses
that spanned the range of adopted 1$\sigma$ confidence intervals
(low$-$high, low$-$low, high$-$low, high$-$high combinations), and
repeats of the nominal configuration for additional randomized initial
angles of the argument of pericenter and mean anomaly. We did not
specifically vary the tidal quantity $k/Q$, as the effect of varying
$k/Q$ is the same as varying the speedup factor since they both
contribute linearly to $\dot{a}$ and $\dot{e}$. We describe our
results in the next subsection.

\subsection{Results: Evolutionary Pathways} \label{mmr_results}

Here we describe the results stemming from our simulations of a 3:1
resonant passage between Remus and Romulus. These results suggest
three evolutionary pathways: capture into resonance with no escape,
temporary capture followed by escape, and no capture. We describe each
of these evolutionary pathways in the following paragraphs.

\subsubsection{Capture with no escape}

\begin{figure}[thbp]
	\centering
	\includegraphics[width=3.3in]{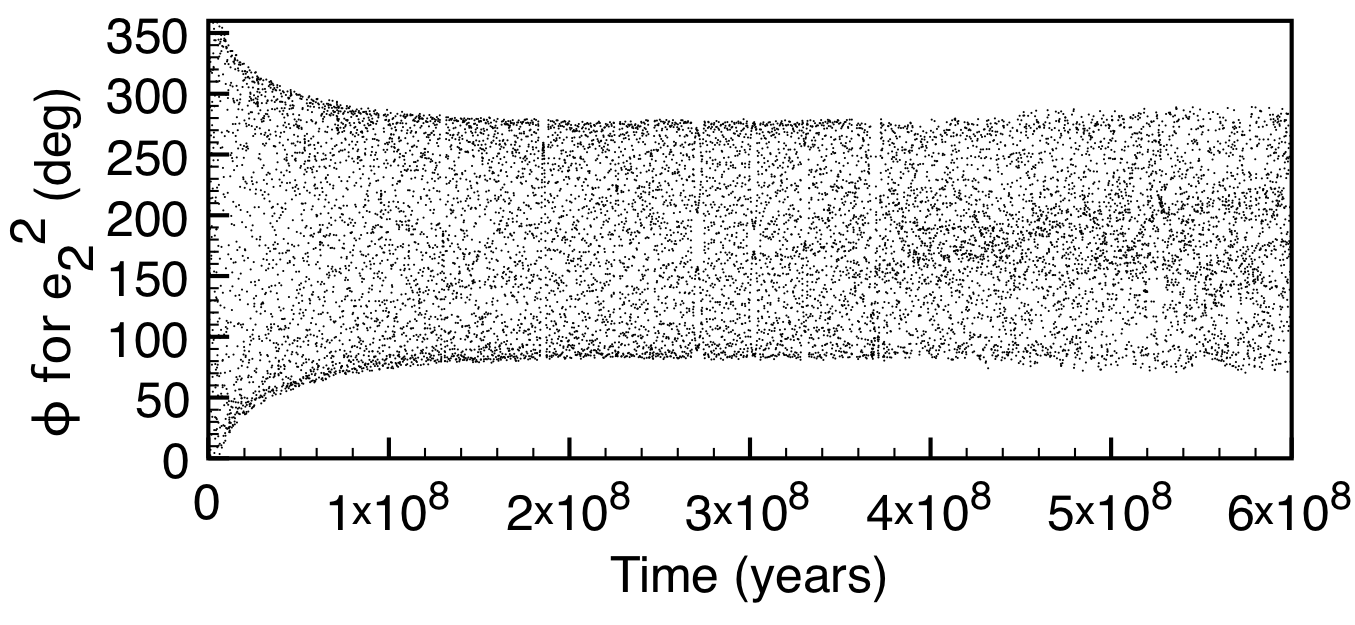}
	\includegraphics[width=3.3in]{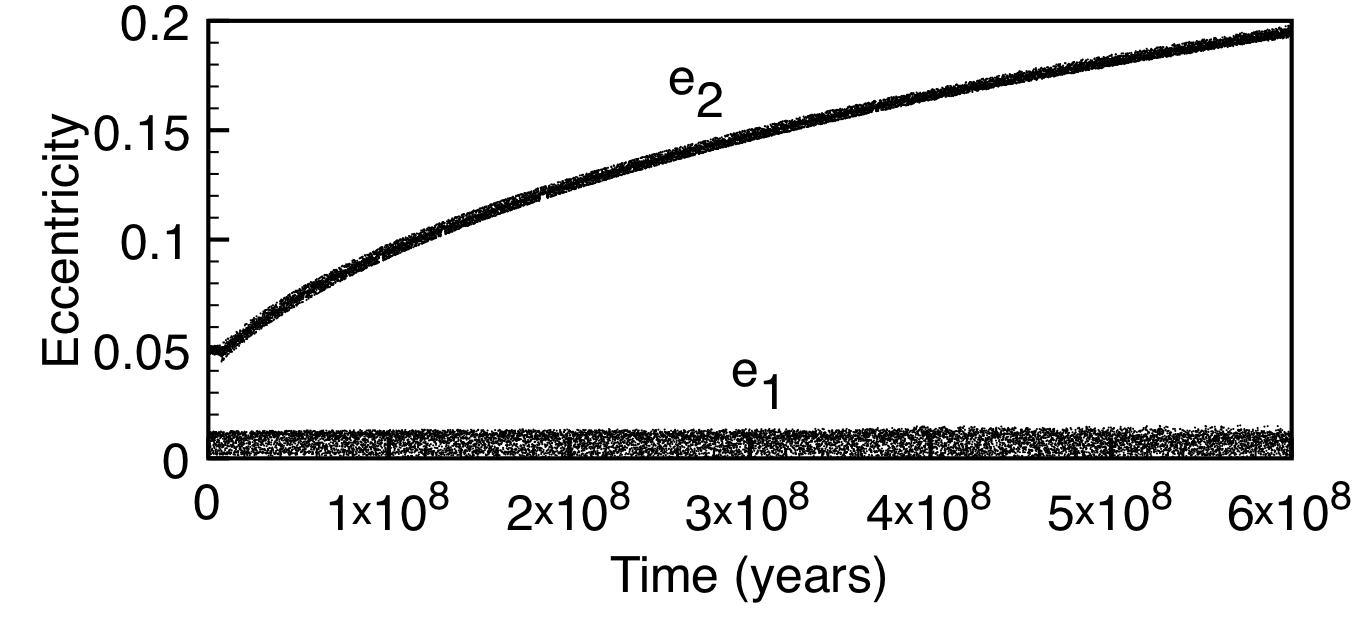}

	\caption{Example of a simulation with resonant capture into
          $e_2^2$ and no escape. Dots show results from numerical
          simulations for the libration angle $\phi$ and eccentricity
          $e$ as a function of time. Subscripts $1$ and $2$ represent
          the inner and outer satellites, respectively.  The
          simulation was started at $a_1/a_2 = 654.0/1352.5 = 0.4835$
          and reached the $e_2^2$ resonance at $a_1/a_2 = 655.2/1352.5
          = 0.4844$, roughly 8 Myr after the start of the simulation.
          The outer satellite's eccentricity will continue to increase
          due to the resonance effects, but the tidal damping effects
          will increase as the eccentricity grows, such that an
          equilibrium value for $e$ may be reached.
\label{mmr_plot1}}

\end{figure}

This scenario, where resonant capture occurs with no escape during the
1 Gyr evolution, was typically observed for the $e_2^2$ resonance.  An
example of such evolution is shown in Figure \ref{mmr_plot1}. Given
that (a) the satellites are not currently observed in the 3:1
resonance and (b) these simulations show no escape from such resonant
capture within a reasonable system lifetime of 1 Gyr, we conclude that
this evolutionary pathway did not occur and we do not discuss it
further.

% %%%%%%%%%%%%%%%%%%%%%%%%%%%%%%%%%%%
% %%%%%%%%%%%%%%%%%%%%%%%%%%%%%%%%%%% 

\subsubsection{Temporary capture followed by escape}

\begin{figure*}[htbp]
	\centering
	\mbox{\subfigure{\includegraphics[width=3.3in]{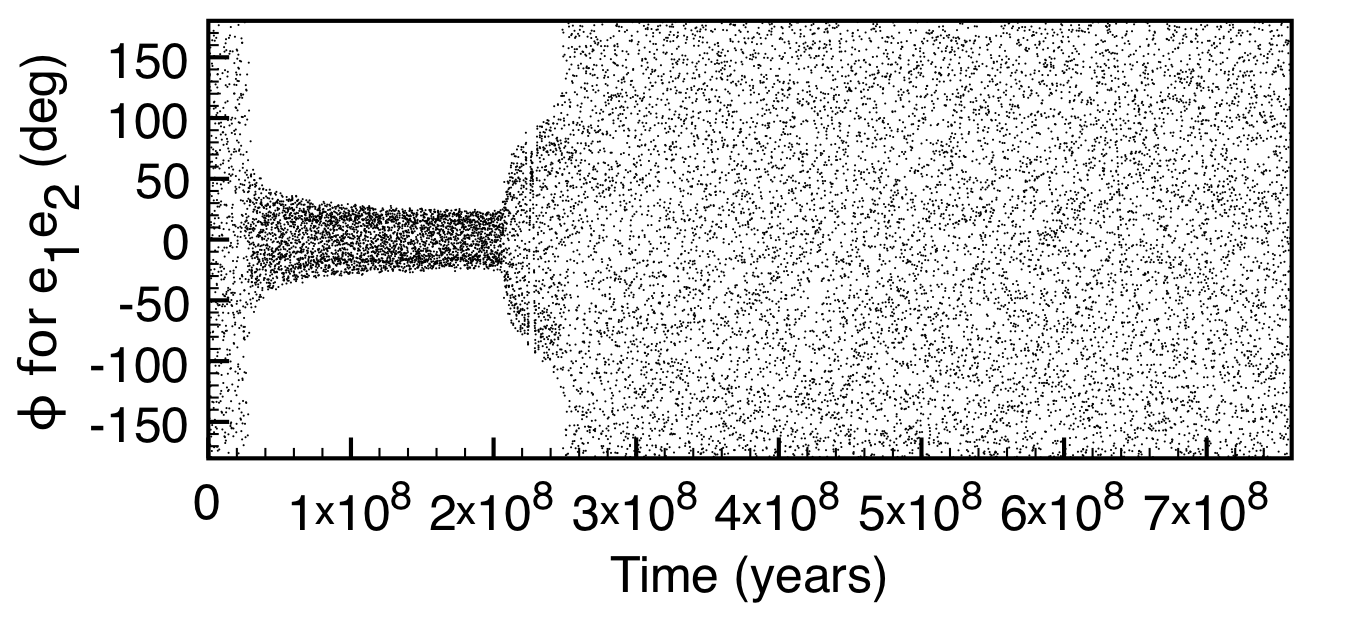}}\quad
	\subfigure{\includegraphics[width=3.3in]{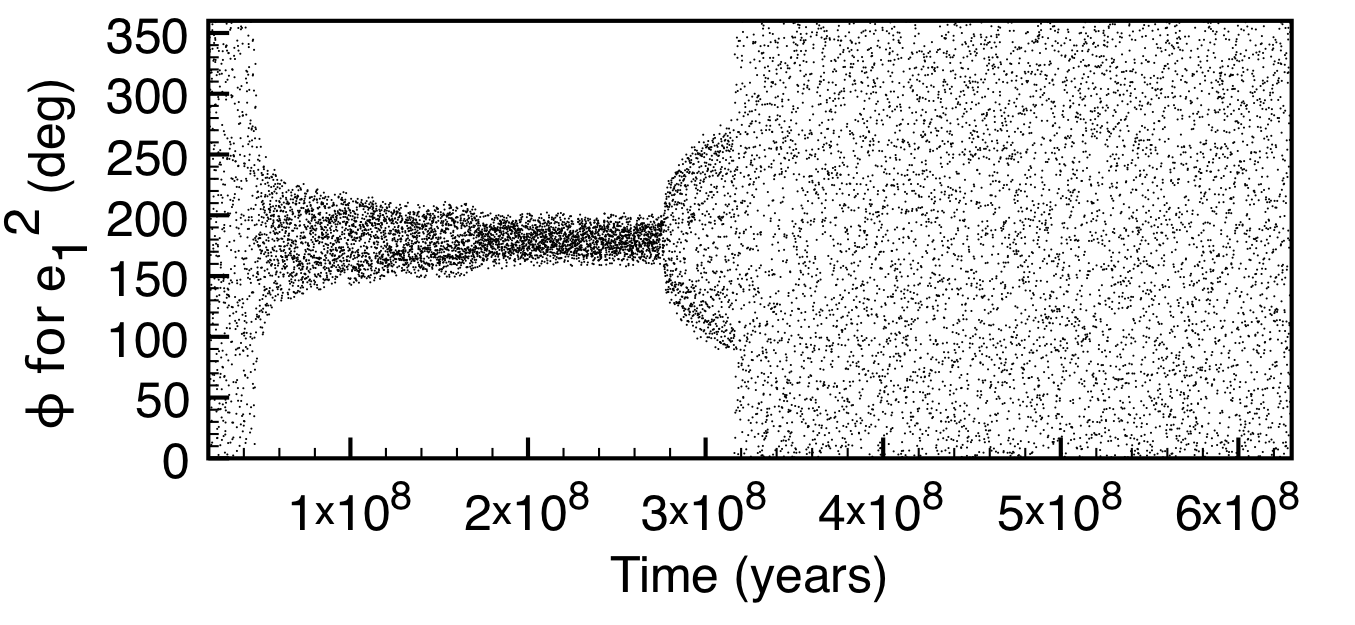}}}
	\mbox{\subfigure{\includegraphics[width=3.3in]{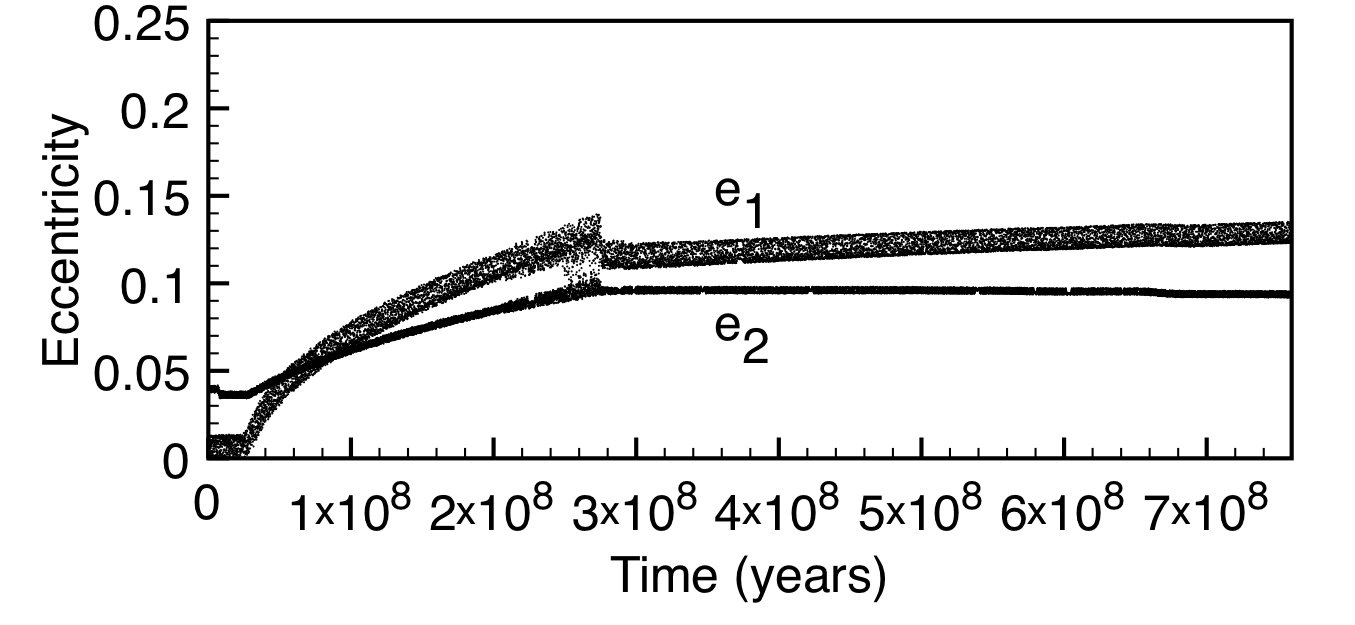}}\quad
	\subfigure{\includegraphics[width=3.3in]{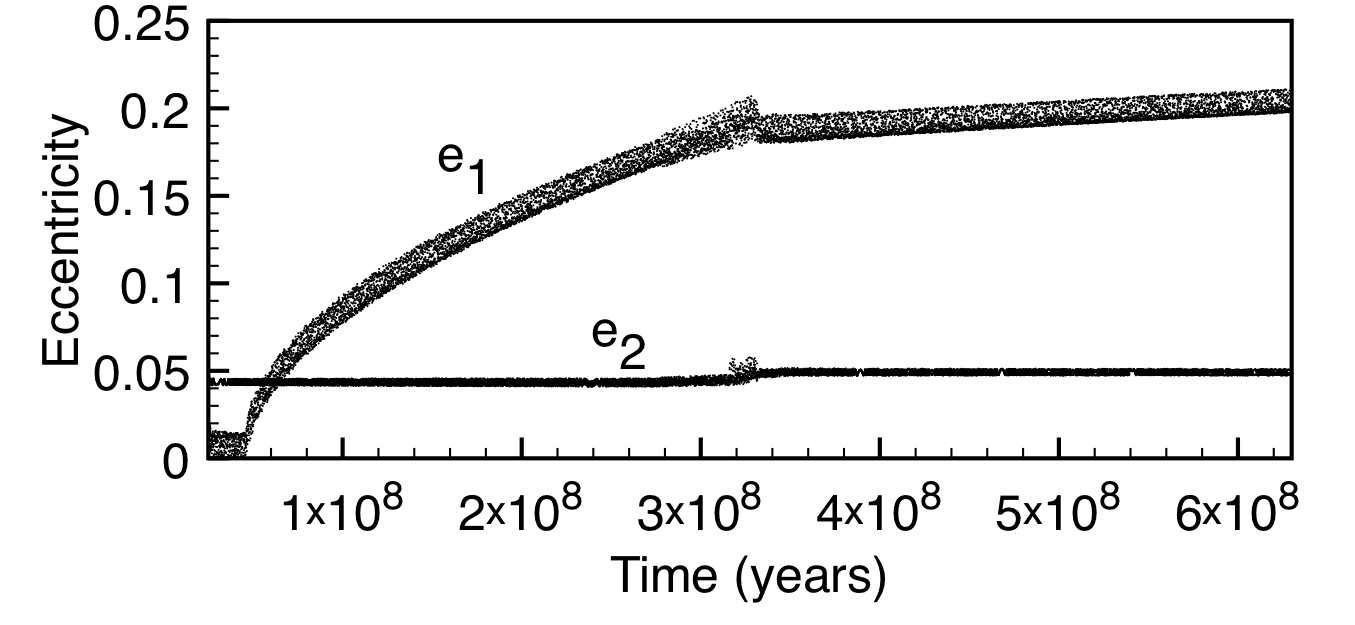}}}
	\caption{Examples of simulations with temporary resonant
          capture followed by escape.  Dots show results from
          numerical simulations for the libration angle $\phi$ and
          eccentricity $e$ as a function of time. Subscripts $1$ and
          $2$ represent the inner and outer satellites, respectively.
          {\bf Left-side plots:} Results of a simulation with
          temporary capture into $e_1e_2$ resonance, where both $e_1$
          and $e_2$ increase. Initial conditions for eccentricity are
          $e_1=0.001$ and $e_2=0.04$.  {\bf Right-side plots:} Results
          of a simulation with temporary capture into $e_1^2$
          resonance, where only $e_1$ increases. Initial conditions
          for eccentricity are $e_1=0.007$ and $e_2=0.04$.
 \label{mmr_plot2}}
\end{figure*}

In this event, resonant capture occurs and is followed by eventual
escape due to growth of the resonant argument. This was a common
outcome for each of the 3 types of resonances. Examples of such
evolution are shown in Figure \ref{mmr_plot2}. Final eccentricities at
the end of our simulations ranged from their initial values up to
$\sim$0.3.  We are unable to place lower bounds on the final
eccentricities because we cannot assess how long the satellites may
have been captured in the resonance.  If high eccentricities resulted
from temporary resonance capture, then sufficient eccentricity
damping may have subsequently occurred to bring high post-resonance
eccentricities to low observed eccentricities.

We investigated whether such eccentricity damping was possible within
a conservative timeframe of 1 Gyr. To do so, we integrated the
$\dot{a}$ and $\dot{e}$ tidal expressions in Equations (\ref{dadt})
and (\ref{dedt}) and considered all Love number models (Equations
\ref{monogold}$-$\ref{rubbjaco}), various post-resonance
eccentricities up to 0.25, and tidal dissipation $Q$ values
(10$-$1000). We assumed that both satellites had densities equal to
that of the primary (1.29 g cm$^{-3}$).

For Remus, we find that eccentricity damping to observed values is
only possible if we assume rubble pile Love number models (either $k
\propto R$ or $k \propto 1/R$) for Remus (there is no restriction on
the primary).  If we make this assumption, damping to observed values
is possible by adopting reasonable values of $Q_p=100$ and
$Q_s=10-100$.  When we assume that Remus is monolithic, even when we
adopt very favorable conditions for eccentricity
damping\footnote[3]{$Q_p=1000$ and $Q_s=10$. Inspection of Equation
(\ref{dedt}) shows that damping can be speeded up by making $Q_p/Q_s$
as large as possible.}, tidal damping to its observed value is only
possible if we assume a post-resonance eccentricity of $\sim$0.032 or
less.  These calculations suggest that if its post-resonance
eccentricity exceeded $\sim$0.032, it is likely that Remus may have an
interior structure more akin to a rubble pile aggregate than a
monolithic body.

For Romulus, damping to observed eccentricities is possible only if
the eccentricity was barely affected while in the resonance (as well
as assuming favorable dissipations conditions: $Q_p=1000$ and
$Q_s=10$).  If the eccentricity reached even modest values
($\sim0.023$) we find that none of the Love number models and
reasonable $Q=10-1000$ values can damp eccentricities to even the
highest possible observed eccentricity (0.011) allowed by our fit
uncertainties.  Therefore, if temporary capture in the 3:1 occurred,
it must not have lasted long enough for the eccentricity of Romulus to
reach values of $\sim 0.023$.  While such a scenario does not entirely
rule out the $e_2^2$ and $e_1e_2$ resonances, it does seem to place
bounds on the acceptable increase in eccentricity due to the 3:1
resonance.

% %%%%%%%%%%%%%%%%%%%%%%%%%%%%%%%%%%%
% %%%%%%%%%%%%%%%%%%%%%%%%%%%%%%%%%%%

\subsubsection{No capture}

In this case, all eccentricity-type resonances are
encountered and none result in capture. For orbits that are slowly
converging toward each other, capture is possible depending on their 
initial eccentricities. When the pre-encounter 
eccentricity is below a critical eccentricity, capture is guaranteed. 
When the pre-encounter eccentricity exceeds a critical eccentricity, 
capture becomes a probabilistic event. For the 3:1 resonance, critical
eccentricities can be estimated as \citep{murr99}
\begin{align} \label{ecrit1}
	e_{\rm 1,crit} = \left[ \dfrac{3}{32f_1} \left( 3^{2/3} \dfrac{M_p}{M_2} + 3^{4/3} \dfrac{M_1}{M_2} \dfrac{M_p}{M_2} \right) \right]^{-1/2},
\end{align}
\begin{align} \label{ecrit2}
	e_{\rm 2,crit} = \left[ \dfrac{3}{32f_2} \left( 3^{2/3}\dfrac{M_2}{M_1}\dfrac{M_p}{M_1} + 9\dfrac{M_p}{M_1} \right) \right]^{-1/2},
\end{align}
where subscripts $p$, $1$, and $2$ represent the primary, Remus, and
Romulus, respectively. The $f_1$ and $f_2$ terms represent functions
of Laplace coefficients $b_{1/2}^{(j)}(\alpha)$. They are \citep[e.g., see
][]{murr99}
\begin{align} \label{f1}
	f_1 = \dfrac{1}{8}(-5j + 4j^2 - 2\alpha D + 4j\alpha D + \alpha^2 D^2)b_{1/2}^{(j)}(\alpha),
\end{align}
\begin{align} \label{f2}
	f_2 = \dfrac{1}{8}(2 - 7j + 4j^2 - 2\alpha D + 4j \alpha D + \alpha^2 D^2) b_{1/2}^{(j-2)}(\alpha) - \dfrac{27}{8}\alpha,
\end{align}
where $j=3$ for the 3:1 resonance, $\alpha=a_1/a_2$ is the ratio of
semi-major axes, and $D=d/d\alpha$ is the differential operator. The
quantity $-27\alpha/8$ in Equation (\ref{f2}) is the indirect term for
the case when $M_2>M_1$. 

Using Equations (\ref{ecrit1}) and (\ref{ecrit2}), we calculate the
critical eccentricities to be $e_{\rm 1,crit} = $ 0.00864 and $e_{\rm
2,crit} = $ 0.00410. Even for initial $e_1$ values that are low
(e.g.,~osculating value of 0.001), $e_1 < e_{\rm 1,crit}$ will not
always be satisfied because short-term eccentricity fluctuations due
to primary $J_2$ will inflate $e_1$ excursions up to $\sim$0.014
(assuming $a_1 = 654$ km in Equation (\ref{j2e}) of Section
\ref{evolution}). For Romulus, marooned farther from the primary such
that $J_2$ effects are lessened, if the pre-encounter eccentricity is
low enough then it is possible that the eccentriciy will always remain
less than the critical eccentricity.  These analytical arguments are
in agreement with the results from our numerical experiments.

If we contemplate scenarios in which resonant capture never occurred
in Sylvia's past, then we must adopt the critical eccentricities as
lower limits on the past eccentricities of Remus and Romulus.  Their
past eccentricities cannot be lower than these limits because
otherwise capture would have been guaranteed. We note that these lower
limit eccentricities are lower than the nominal observed
eccentricities (Table \ref{bestfit}), and hence this evolutionary
pathway is a plausible scenario without requiring any significant
modifications in eccentricity over time.

To summarize these results, from these 3 pathways we find that both
(a) temporary capture followed by escape and (b) no capture are
plausible scenarios that occurred when Remus and Romulus encountered
the 3:1 resonance. If pathway (a) occurred, our calculations of the
necessary damping required to bring post-resonance eccentricities to
observed values show this is possible for Remus but may be
prohibitively long for Romulus, depending on its post-resonance
eccentricity.  Therefore it is unlikely that a substantial increase in
the eccentricity of Romulus occurred, even if the system was
temporarily captured in the $e_2^2$ or $e_1e_2$ resonances.  If
pathway (b) occurred, we can set lower limits on past eccentricities
of both satellites to be equal to their critical eccentricities.

% %%%%%%%%%%%%%%%%%%%%%%%%%%%%%%%%%%%
\section{Conclusions} \label{conclusion}

The goals of this study were to characterize Sylvia's current orbital
configuration and masses as well as to illuminate the past orbital
evolution of this system. Our work can be summarized as follows:

(1) We reported new astrometric observations of Sylvia in 2011 that
increased the number of existing epochs of astrometry by over 50\%.
These new observations extended the existing baseline of observations
to 7 years (for Remus) and to 10 years (for Romulus).

(2) We fit a fully dynamical 3-body model to the available astrometric
data. This model simultaneously solved for orbits of both satellites,
individual masses, and the primary's oblateness (Table
\ref{bestfit}). We found that the primary has a density of
1.29$\pm$0.39 g cm$^{-3}$ and is oblate with a $J_2$ value in the
range of 0.0985$-$0.1.  Constraints on satellite radii can be obtained
from the mass determinations by assuming that the satellites have a
bulk density equal to that of the primary; we find $\sim$4.5$-$6.1 km
for Remus and $\sim$2.6$-$8.2 km for Romulus.  These ranges would have
to be modified if the actual density of the primary or of the
satellites was different from the nominal value assumed here. The
orbits of the satellites are relatively circular. We find that the
primary's spin pole is best fit when aligned to Romulus' orbital pole,
and that the satellites' orbit poles are coplanar to within one
degree.

(3) We numerically investigated the short-term and long-term stability
of the orbits of Sylvia's satellites.  There are periodic fluctuations
in eccentricity for both satellites, most notably for the inner
satellite Remus. We verified that these eccentricity excursions are
due to the effects of primary oblateness.  From long-term integrations
we found that the system is in a very stable configuration, in
agreement with previous investigations.

(4) We studied the past orbital evolution of Sylvia's satellites,
including the most recent low-order MMR resonance crossing, which is
the 3:1.  We used direct N-body integrations with forced tidal
migration to model such an encounter.  To examine the case of resonant
capture followed by escape, we calculate the tidal damping timescale
to go from the post-encounter eccentricity to the observed value.
Using available tidal models, we find that the damping timescale for
Romulus can be prohibitively large if its post-resonance eccentricity
exceeded $\sim$0.023.  This suggests that the system crossed the
$e_2^2$ and $e_1e_2$ resonances without capture, or that it was not
captured in these resonances for a sufficient duration to
substantially increase the eccentricity of Romulus.  Similar timescale
constraints from tidal damping also imply that Remus may have a rubble
pile structure if its post-resonance eccentricity exceeded
$\sim$0.032.  Alternatively, if no capture in any resonance occurred
then we are able set lower limits on their past eccentricities
($e_1=0.00864$ and $e_2=0.00410$).

The detailed characterization of Sylvia presented in this paper has
allowed for analyses of its orbital evolution. Such studies of triple
systems are important in order to understand their key physical
properties, orbital architectures, and intriguing evolutionary
histories.

\acknowledgments

We thank Stan Peale, Man Hoi Lee, and Fred Davies for useful 
discussions. We are grateful to the reviewer for helpful 
comments.

Some of the data presented herein were obtained at the W.M. Keck
Observatory, which is operated as a scientific partnership among the
California Institute of Technology, the University of California and
the National Aeronautics and Space Administration. The Observatory was
made possible by the generous financial support of the W.M. Keck
Foundation.  The authors wish to recognize and acknowledge the very
significant cultural role and reverence that the summit of Mauna Kea
has always had within the indigenous Hawaiian community.  We are most
fortunate to have the opportunity to conduct observations from this
mountain.  This work was based in part on observations made with ESO
telescopes at the La Silla Paranal Observatory under programme ID
088.C-0528.  This research was done using resources provided by the
Open Science Grid (OSG), which is supported by the National Science
Foundation and the U.S. Department of Energy's Office of Science.  We
thank Mats Rynge for his assistance with the OSG.

This work was partially supported by NASA Planetary Astronomy grant
NNX09AQ68G.

% %%%%%%%%%%%%%%%%%%%%%%%%%%%%%%%%%%%
\bibliographystyle{apj}
\bibliography{sylvia}

\end{document}